\documentclass[twocolumn,times]{aastex62}

\newcommand{\Gaia}{\textit{Gaia }}
\newcommand{\unit}[2]{\ensuremath{\textrm{#1}^{#2}}}
\newcommand{\sub}[2]{\ensuremath{{#1}_{\mathrm{#2}}}}
\newcommand{\super}[2]{\ensuremath{{#1}^{\mathrm{#2}}}}
\newcommand{\supersub}[3]{\ensuremath{{#1}^{\mathrm{#2}}_{\mathrm{#3}}}}

\graphicspath{{./}{figures/}}

\received{}
\revised{}
\accepted{}
\submitjournal{ApJ}

\shorttitle{Cosmological mock Gaia surveys}
\shortauthors{Sanderson et al.}
\newcommand{\Msun}{\,\textnormal{M}_\odot}

\newcommand{\kpc}{\,\textnormal{kpc}}
\newcommand{\pc}{\,\textnormal{pc}}

\newcommand{\cci}{\,\textnormal{cm} ^ {-3}}
\newcommand{\Gyr}{\,\textnormal{Gyr}}

\newcommand{\K}{\,\textnormal{K}}

\newcommand{\Mthm}{M_{200{\rm m}}}
\newcommand{\Rthm}{R_{200{\rm m}}}

\newcommand{\upenn}{Department of Physics \& Astronomy, University of Pennsylvania, 209 S 33rd St., Philadelphia, PA 19104, USA}
\newcommand{\flatiron}{Center for Computational Astrophysics, Flatiron Institute, 162 5th Ave., New York, NY 10010, USA}
\newcommand{\caltech}{TAPIR, Mailcode 350-17, California Institute of Technology, Pasadena, CA 91125, USA}
\newcommand{\davis}{Department of Physics, University of California, Davis, CA 95616, USA}
\newcommand{\ucsd}{Department of Physics, Center for Astrophysics and Space Science, University of California at San Diego, 9500 Gilman Drive, La Jolla, CA 92093, USA}
\newcommand{\nw}{Department of Physics and Astronomy and CIERA, Northwestern University, 2145 Sheridan Road, Evanston, IL 60208, USA}
\newcommand{\berkeley}{Department of Astronomy and Theoretical Astrophysics Center, University of California Berkeley, Berkeley, CA 94720}
\newcommand{\sydney}{Sydney Institute for Astronomy, School of Physics, University of Sydney, NSW 2006, Australia}

\begin{document}

\title{Synthetic Gaia surveys from the FIRE cosmological simulations of Milky-Way-mass galaxies}

\correspondingauthor{Robyn Sanderson}
\email{robynes@sas.upenn.edu}

\author[0000-0003-3939-3297]{Robyn E. Sanderson}
\altaffiliation{NSF Astronomy \& Astrophysics Postdoctoral Fellow}
\affiliation{\caltech}
\affiliation{\upenn}
\affiliation{\flatiron}

\author{Andrew Wetzel}
\affiliation{\davis}

\author{Sarah Loebman}
\altaffiliation{Hubble Fellow}
\affiliation{\davis}

\author{Sanjib Sharma}
\affiliation{\sydney}

\author{Philip F. Hopkins}
\affiliation{\caltech}

\author{Shea Garrison-Kimmel}
\affiliation{\caltech}

\author{Claude-Andr\'e Faucher-Gigu\`ere}
\affiliation{\nw}

\author{Du\v san Kere\v s}
\affiliation{\ucsd}

\author{Eliot Quataert}
\affiliation{\berkeley}

\begin{abstract}

With \Gaia Data Release 2, the astronomical community is entering a new era of multidimensional surveys of the Milky Way. This new phase-space view of our Galaxy demands new tools for comparing observations to simulations of Milky-Way-mass galaxies in a cosmological context, to test the physics of both dark matter and galaxy formation.
We present \textsf{ananke}, a framework for generating synthetic phase-space surveys from high-resolution baryonic simulations, and use it to generate a suite of synthetic surveys resembling \Gaia DR2 in data structure, magnitude limits, and observational errors.
We use three cosmological simulations of Milky-Way-mass galaxies from the \textit{Latte} suite of the Feedback In Realistic Environments (FIRE) project, which feature self-consistent clustering of star formation in dense molecular clouds and thin stellar/gaseous disks in live cosmological halos with satellite dwarf galaxies and stellar halos. We select three solar viewpoints from each simulation to generate nine synthetic Gaia-like surveys. We sample synthetic stars by assuming each star particle (of mass $7070 \Msun$) represents a single stellar population. At each viewpoint, we compute dust extinction from the simulated gas metallicity distribution and apply a simple error model to produce a synthetic \Gaia-like survey that includes both observational properties and a pointer to the generating star particle. We provide the complete simulation snapshot at $z = 0$ for each simulated galaxy. We describe data access points, the data model, and plans for future upgrades. These synthetic surveys provide a tool for the scientific community to test analysis methods and interpret \Gaia data.

\end{abstract}


\section{Introduction} \label{sec:intro}

A new generation of observational projects is poised to revolutionize our understanding of resolved stellar populations of the Milky Way (MW) and MW-mass galaxies at an unprecedented level of detail, ushering in an era of precision studies of galaxy formation. In the MW itself, astrometric, spectroscopic, and photometric surveys will measure three-dimensional positions and velocities and numerous elemental abundances for stars from the disk to the halo, as well as for many satellite dwarf galaxies. In the Local Group and beyond, the Hubble Space Telescope (HST), James Webb Space Telescope (JWST) and eventually the Wide-Field Infrared Survey Telescope (WFIRST) will deliver pristine views of resolved stellar populations. The groundbreaking scale and dimensionality of this new view of resolved stellar populations in galaxies challenge us to develop new theoretical tools to robustly compare these surveys to simulated galaxies, in order to take full advantage of our new ability to make detailed predictions for stellar populations within a cosmological context.

Broadly speaking, two classes of modeling tools exist for generating synthetic stellar observations of the MW. One class of tools samples empirically derived density distributions generally assumed to be in dynamical equilibrium, such as the Besan\c{c}on model \citep{Robin2003}, galfast \citep{Juric2010}, TRILEGAL \citep{trilegal}, and GalMod \citep{pasetto18}. The Besan\c{c}on model in particular has been crucial to forecasting and preparation for \Gaia \citep[e.g.][]{2012A&A...543A.100R}; a \Gaia DR2-like catalog based on this model was recently released by \citet{2018PASP..130g4101R}. The other common approach is to re-sample star particles from cosmological simulations of galaxy formation in a manner that preserves phase-space properties, such as the Galaxia\footnote{http://galaxia.sourceforge.net} implementation of the \citet{bj05} mock stellar halos by \citet{sharma11} or the resampling of cosmological simulations in \citet{Lowing2015} and later \citet{Grand2018}.
In the latter case, the phase space density usually is computed using a strategy such as that employed in the publicly available code EnBiD\footnote{\url{https://sourceforge.net/projects/enbid/}} \citep{2006MNRAS.373.1293S}, which partitions particles and chooses an adaptive smoothing length based on the local Shannon entropy in each dimension, avoiding the need to define a metric for the six-dimensional distance.

While the empirical, equilibrium approach has the advantage of being tied closely to true MW data, it has the disadvantage of requiring a simple analytic, generally equilibrium description of structure, lacking the rich complexity of non-axisymmetric structures like spiral arms in the disk, over-densities in the stellar halo, satellite galaxies, and dark-matter substructure. The $N$-body resampling approach from cosmological simulations has the disadvantage of not generating a one-to-one match of the MW; however, catalogs generated from $N$-body simulations can capture the non-equilibrium structures born out of a fully cosmological context.

Synthetic surveys generated from such cosmological simulations can serve as valuable aids in characterizing the efficacy of analysis tools to recover ``ground truth,'' from characterizing the underlying potential and dark-matter distribution to recovering the evolutionary history of a simulated galaxy.
Thus, while a cosmologically simulated galaxy will not replicate the MW perfectly, it can provide great value as a testbed for discovery and recovery of underlying structural, dynamic, and abundance trends.
For example, \citet{Grand2018} recently published two sets of mock \Gaia DR2 catalogs generated from 6 cosmological high-resolution MW-mass simulations from the AURIGA project.
One set of the mock catalogs were generated using the public SNAPDRAGONS code \citep{Hunt2015} and the other used a phase-space smoothing kernel as described in \citet{Lowing2015}.
\citet{Grand2018} provides a powerful demonstration of how properties of the stellar halo and young stellar disk can be recovered from these catalogs.

%The Besan\c{c}on model reproduces the stellar content of the Galaxy, assuming stars belong to one of several distinct structural components that can be modeled with a smooth distribution. The modeling of each population is based on a set of evolutionary tracks, assumptions on density distributions, constrained either by dynamical considerations or by empirical data, and guided by a scenario of formation and evolution.

However, existing synthetic surveys of $N$-body simulations still suffer from some limitations. Previous semi-analytic studies, such as the resampling of \citet{bj05} in \citet{sharma11} or the resampling of \citet{cooper10} in \citet{Lowing2015}, are limited by the use of dark-matter-only cosmological simulations as the basis for generating the stellar phase-space distributions. This prevents a self-consistent realization of both the disk-like central galaxy and the accreted halo, and relies on a prescription for translating the phase-space distribution of the dark matter into that of the stars. Such prescriptions can be quite sophisticated, as is the case with \citet{cooper10}, but cannot self-consistently capture the ongoing interaction between star formation, feedback processes, and the dark matter halos which is of especial importance in the smaller accreted galaxies that eventually form the stellar halo. Neither can these simulations more than approximately account for the contribution of the central galaxy's disk to the tidal destruction of accreted galaxies, which has been shown to be significant especially within the range of \Gaia \citep[e.g.][]{GarrisonKimmel2017}.

Another limitation of attempts to resample the stellar distributions of cosmological-hydrodynamical simulations, as in \citet{Grand2018}, is the choice to use the observed MW extinction map in the synthetic surveys, rather than determining the extinction from the metal-enriched gas distribution in the simulated galaxy. This choice can produce both large- and small-scale discrepancies in the stellar density distribution predicted by the synthetic survey: on small scales, the extinction should be correlated with regions where young stars are forming, while on large scales it should be related to the height of the young thin disk and correlated with any spiral features that exist in the simulated galaxy. 

To address these limitations, we use the \textit{Latte} suite of simulations of Milky-Way-mass galaxies \citep{Wetzel2016, Hopkins2017fire}, run as part of the Feedback In Realistic Environments (FIRE) simulation project.\footnote{FIRE project website: \url{http://fire.northwestern.edu}}
Using these simulations as the basis for generating synthetic surveys allows us to incorporate the effects of baryonic processes in a cosmological context, improving on semi-analytic analyses of dark-matter-only simulations while retaining sufficient resolution to include kinematically cold, single-age, single-metallicity stellar populations, at mass resolution ($7070 \Msun$) comparable to the masses of (massive) star clusters.
Given their spatial resolution (as small as $1 \pc$ in gas) and tracking of cold (down to $10 \K$) gas, the simulations start to resolve the formation of giant molecular clouds and thus the clustered formation of young stars.
These simulations also self-consistently track the metal enrichment of gas, permitting us to encapsulate all the correlations between (young) stars and gas, including the extinction near star-forming regions, assuming a fixed ratio of dust to metal-enriched gas. The result is an extincted synthetic survey of the simulated galaxy that leaves intact important observational relationships between gas, extinction, and stellar populations, as well as the important theoretical relationships between cosmology (dark matter) and galaxy formation.

In this paper we describe a new framework, \textsf{ananke}, for generating realistic synthetic star catalogs and mock stellar surveys from cosmological baryonic simulations, and we present a set of synthetic phase-space surveys created from the textit{Latte} FIRE-2 simulations that are designed to resemble Data Release 2 of the \Gaia astrometric survey (Gaia DR2, \citealt{DR2}). We name this framework for generating synthetic surveys \textsf{ananke}\footnote{In Greek mythology, Ananke is the primordial goddess of necessity and inevitability; along with Chronos, she marks the beginning of the cosmos. In some versions of Greek mythology, she created Gaia.}.

We describe the underlying simulations in \S \ref{sec:sims}, the assumptions used to define the solar viewpoint in \S \ref{sec:coords}, the process for creating synthetic stars from star particles in \S \ref{sec:mock-catalog}, and the extinction and error models used to create the final synthetic surveys in \S \ref{sec:synthetic-survey}. In \S \ref{sec:results} we present preliminary characterizations of the surveys and describe the data model and access modes for the public versions. In \S \ref{sec:quickstart} we provide some guidelines for new users of the surveys, and in \S \ref{sec:future} we discuss a few of the many uses of this new resource.

\section{Simulations}
\label{sec:sims}

\subsection{GIZMO code and FIRE-2 model}
\label{subsec:numerics}

Cosmological ``zoom-in' simulations, which model a selected region at high resolution embedded within a lower-resolution cosmological background \citep[e.g.,][]{kw93, Onorbe2014}, now achieve sufficient dynamic range to resolve individual star-forming regions within galaxies, allowing the formation of realistic stellar populations that can connect with detailed observations of the MW.
Such cosmological simulations allow one to examine realistic formation histories of MW-like systems---with cosmic accretion, galactic outflows, time-dependent asymmetric gravitational potentials, and orbiting satellites---enabling the study of the entire MW system within one simulation, at a resolution necessary for detailed stellar modeling.

\citet{Hopkins2017fire} provides all details of our simulation methodology; we briefly describe the most important aspects here.

We use three cosmological zoom-in simulations of individual MW-like galaxies from the \textit{Latte} suite \citep{Wetzel2016} of FIRE-2 simulations.
We ran these simulations using \textsc{GIZMO}\footnote{A public version of \textsc{GIZMO} is available at: \url{http://www.tapir.caltech.edu/~phopkins/Site/GIZMO.html}} \citep{Hopkins2015}, a multi-method gravity plus hydrodynamics code.
The hydrodynamics are solved using the meshless finite-mass (``MFM'') method, a mesh-free Lagrangian finite-volume Godunov method that automatically provides adaptive spatial resolution while maintaining conservation of mass, energy, momentum, and angular momentum.
Gravity is solved with an improved version of the Tree-PM solver from GADGET-3 \citep{Springel2005gadget2}, using fully adaptive and fully conservative gravitational force softenings for gas \citep[see][]{Hopkins2015}, matching the hydrodynamic resolution.

Our simulations were run with the FIRE-2 physics models from \citet{Hopkins2017fire}. FIRE-2 incorporate radiative cooling and heating from $10 - 10^{10}$ K, including free-free, photo-ionization and recombination, Compton, photoelectric and dust collisional, cosmic ray, molecular, metal-line, and fine-structure processes, explicitly accounting for 11 elements (H, He, C, N, O, Ne, Mg, Si, S, Ca, Fe).
This includes photo-ionization/heating from a redshift-dependent, spatially uniform ultraviolet background, including cosmic reionization, from \citet{FaucherGiguere2009}, and an approximate model for local sources and self-shielding. The simulations achieve sufficiently high dynamic range to resolve phase structure of the inter-stellar medium (ISM), allowing gas to condense into resolved giant molecular clouds (\citealt{Hopkins2017fire}, Lakhlani et al., in prep.).

Star formation occurs only in \textit{self-gravitating} gas \citep[following][]{Hopkins13} that also is molecular and self-shielding \citep[following][]{KrumholzGnedin2011}, Jeans unstable, and exceeds a minimum density threshold, $n_{\rm SF} > 1000 \cci$.
These star-formation criteria naturally produce clustered stellar populations in these simulations (\citealt{Hopkins13}; Loebman et al., in prep.).
Once a star particle forms, the simulation explicitly follows several stellar feedback mechanisms, including (1) local and long-range momentum flux from radiation pressure (in the initial UV/optical single-scattering, and re-radiated light in the IR), (2) energy, momentum, mass and metal injection from supernovae (core-collapse and Ia), and stellar mass loss (dominated by O,B and AGB stars), and (3) photo-ionization and photo-electric heating.
Every star particle is treated as a single stellar population with known mass, age, and metallicity; all feedback event rates, luminosities and energies, mass-loss rates, and other quantities are tabulated directly from stellar evolution models \citep[STARBURST99 v7.0;][]{Leitherer1999, leitherer14}, assuming a \citet{Kroupa2001} IMF.

Supernovae (core-collapse and Ia) and stellar winds generate and disperse metals, which are then deposited into surrounding gas particles.
We adopt nucleosynthetic yields for supernovae Ia from \citet{Iwamoto1999}, where the rates follow \citet{Mannucci2006}, including both prompt and delayed populations; core-collapse supernovae yields are from \citet{Nomoto2006}; yields from stellar winds (AGB and O/B-stars) are from a compilation of \citet{VanDenHoek1997, Marigo2001, Izzard2004}.
We initialize all gas particles with a metallicity floor of $\left[M_i / H \right] = -4$ in the initial conditions (to prevent numerical problems in cooling).
These simulations also include an explicit treatment for un-resolved turbulent diffusion of metals in gas \citep{Hopkins16turb,Su2017}, which produces more realistic abundance distributions in both the MW-like galaxies (Wetzel et al., in prep.) and in their satellite dwarf galaxies \citep{Escala2018}.

\subsection{Initial Conditions}
\label{subsec:latte}

Our simulations are drawn from a suite of individual MW-mass halos that are simulated with the same resolution, cosmology, and physics model.
We first run a dark matter-only simulation within a periodic volume of length $85.5$ Mpc with $\Lambda$CDM cosmology: $\Omega_\Lambda = 0.728$, $\Omega_{\rm matter} = 0.272$, $\Omega_{\rm baryon} = 0.0455$, $h = 0.702$, $\sigma_8 = 0.807$, and $n_s = 0.961$.
From this, we select halos at $z = 0$ based \textit{only} on their mass, $\Mthm = 1 - 2 \times 10^{12} \Msun$, and an isolation criterion (no neighboring halos of similar mass within at least $5 \, \Rthm$) to limit computational cost.
We select a sample of $\sim 10$ halos for simulation \citep[listed in][]{2017arXiv171203966G}, agnostic to any halo properties beyond mass and isolation, including formation history, concentration, spin, or subhalo population.
We then trace particles within $5 \,\Rthm$ back to $z = 99$ and regenerated the encompassing convex hull at high resolution, embedded within the lower-resolution volume, using \textsc{MUSIC} \citep{HahnAbel2011}.
Rerun to $z = 0$, all of the zoom-in regions are uncontaminated with low-resolution dark matter out to at least $d_{\rm host} = 600 \kpc$.
Within the zoom-in region, the particle mass resolution is $m_{\rm dm} = 35,000 \Msun$ and $m_{\rm gas,initial} = m_{\rm star,initial} = 7070 \Msun$ (though because of stellar mass loss, at $z = 0$, a typical star particle has $m_{\rm star} \approx 5000 \Msun$, and individual gas particle masses can be up to $\sim 2-3$ times higher).
Dark matter and stars have fixed gravitational softening: $h_{\rm dm} = 40$ pc and $h_{\rm star} = 4$ pc (Plummer equivalent). Gas particles use an adaptive softening, which avoids artificially imposing a maximum density in gas clouds. The choices and implications of force-resolution and interparticle spacing are discussed extensively in \citet[][see Table 3 and related discussion in the text]{Hopkins2017fire}.
The minimum gas resolution (inter-element spacing) and softening/smoothing length reached in each simulation (in the densest regions) is $\sim 1\,$pc. This minimum is reached in the regions where gas is undergoing star formation, since the criteria for star formation (self-gravitating, self-shielding, Jeans-unstable gas with $n_H > 1000 \unit{cm}{-3}$) are based on the gas density and temperature. The gas in which stars form has a density corresponding to interparticle distances of $\lesssim$ 4 pc, hence the choice of softening length for the star particles. For reference, the median Plummer-equivalent smoothing length in the cold gas ($T < 10^4$ K) in the 3 disks used in this work is $\sim$30 pc. These characteristic softening/smoothing lengths are the relevant ones for understanding the conditions under which the simulated star particles acquire their initial phase-space distribution.

For our initial synthetic catalogs, we select three galaxies (m12i, m12f, and m12m) which are approximately MW-like in terms of stellar and gas mass, size, and stellar morphology.
We consider these three systems the most immediately useful for generating synthetic surveys, though we plan to release synthetic catalogs from all of our MW-mass simulations in the future.
These galaxies first were presented in \citet{Wetzel2016} (m12i), \citet{GarrisonKimmel2017} (m12f), and \citet{Hopkins2017fire} (m12m).
Tables~\ref{tbl:halos} and \ref{tbl:galaxies} list their halo-wide and galaxy-wide properties at $z = 0$.\footnote{Movies showing the formation histories of these galaxies are at: \url{http://www.tapir.caltech.edu/~sheagk/firemovies.html}}
% * <arwetzel@gmail.com> 2018-06-26T00:14:03.806Z:
% 
% change or add to this url a permalink on the FIRE website?
% 
% ^.
Table \ref{tbl:galaxies} compares some global properties of the three simulated galaxies with MW values from \citet{bhg16}.

\begin{figure*}
\centering
\includegraphics[width = 0.4 \textheight]{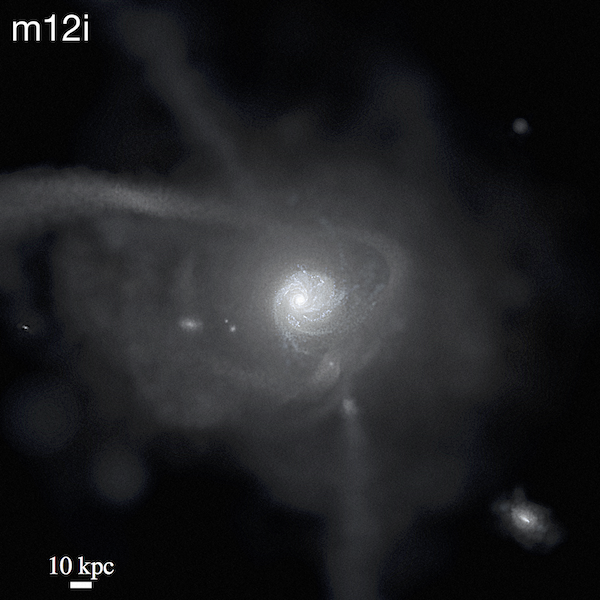} \\
\vspace{2 mm}
\includegraphics[width = 0.6 \textheight]{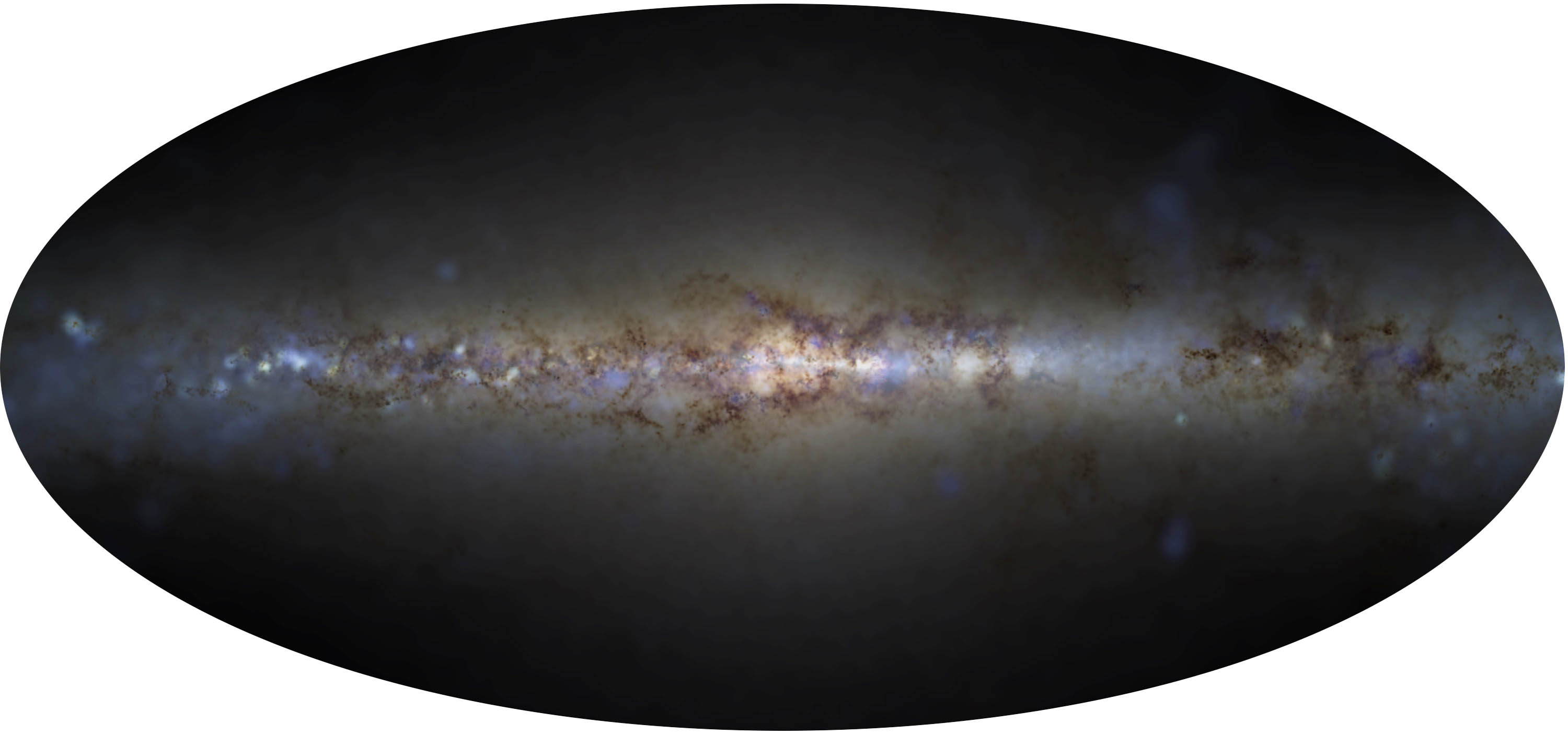} 
\caption{
Mock real-color images, computed using stellar SEDs, of m12i at $z = 0$, as an example of our simulation suite.
\textbf{Top}: Zoom-out image shows a disk-dominated MW-like galaxy at center, a realistic population of satellite dwarf galaxies, and a diffuse stellar halo, including tidal streams and shells from disrupted satellites.
The simulations achieve sufficiently high dynamic range to capture dense star clusters (visible in the central disks and the streams) and satellite dwarf galaxies, all within a live cosmological setting.
\textbf{Bottom}: Mock Galactic (Aitoff) projection, including dust extinction, as seen at the solar circle of m12i.
Individual, filamentary giant molecular cloud (GMC) complexes and young star clusters are visible, and the galaxy has a thin plus thick disk morphology.
}
\label{fig:latte}
\end{figure*}

As an example of this suite, Figure~\ref{fig:latte} shows two images of m12i, demonstrating its ability to simultaneously model the formation of a MW-like stellar disk, a realistic population of satellites, and a realistic stellar halo with streams and shells.

\subsection{Properties of simulated Milky-Way-mass galaxies}

Before creating synthetic surveys it is important to establish that the underlying simulations produce reasonably realistic galaxies. The scheme described in Section \ref{subsec:numerics} does indeed result in galaxies with many properties that reasonably agree with those of the MW, M31, and similar-mass galaxies at $z \sim 0$, without any ``fine-tuning'', including: their stellar-to-halo mass relation \citep{Hopkins2017fire}, stellar thin plus thick disk morphology and metallicity gradients \citep{Ma2017}, $HI$ gas kinematics \citep{ElBadry2018}; giant molecular clouds (Lakhlani et al., in prep.); circum-galactic medium observations of $HI$ and $OVI$ as compared with the COS-Halos survey (Hummels et al., in prep.), realistic populations of satellite dwarf galaxies that do not suffer from the ``missing satellites'' or ``too-big-to-fail'' problems \citep{Wetzel2016, 2018arXiv180604143G} and have realistic metallicity distributions \citep{Escala2018}; and stellar halos \citep{Sanderson2017c, Bonaca2017}.
Using the FIRE-1 simulations, which implemented the same stellar physics (though with somewhat different numerical implementations) and used a SPH hydrodynamics solver, we showed that energy and momentum injection by stellar feedback on the scale of star-forming regions as modeled in FIRE produces a Kennicutt-Schmidt relation \citep{Hopkins2014a,orr18}, galactic winds \citep{muratov15, muratov17, aa17}, and high-redshift circum-galactic medium properties \citep{FaucherGiguere2015, cafg16} in broad agreement with observational constraints, without fine-tuning of parameters to fit these observations.

\begin{figure}
\centering
\includegraphics[width = 0.5 \textwidth]{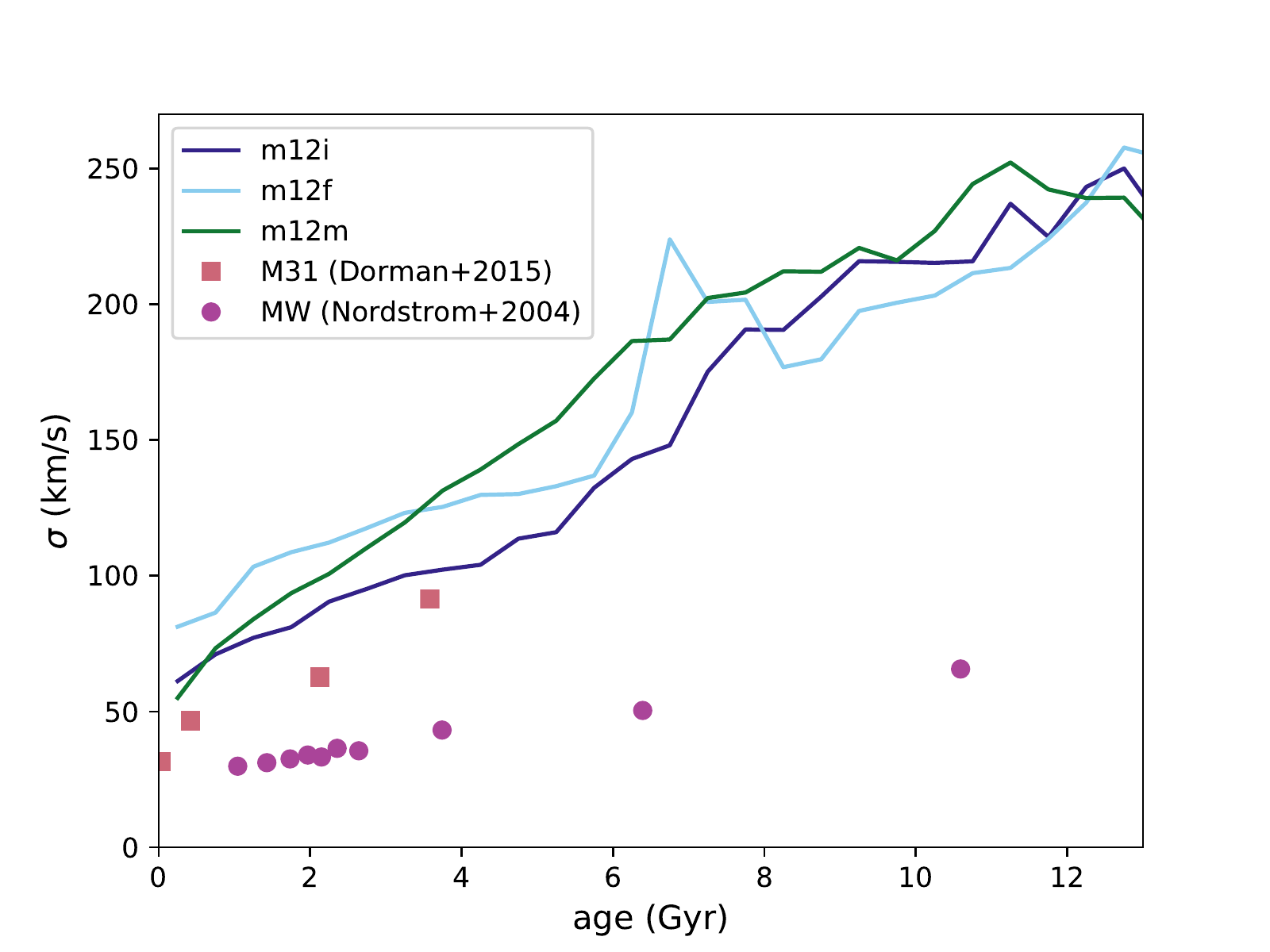} \\
\caption{
Relation between age and total velocity dispersion for star particles within $|z| < 0.3$ kpc, $7.95 < R < 8.45$ kpc, $0^{\circ} \leq \phi < 360^{\circ}$ for m12i, m12f, m12m (shown in dark blue, light blue, and forest green respectively).
Stars in these disks are born hot and are steadily heated with time, similar to M31 \citep[shown in salmon, assuming constant star formation rate,][]{Dorman2015}.
The MW relation \citep[shown in magenta,][]{Nordstrom2004} is substantially cooler and possibly a kinematic outlier.
}
\label{fig:sigma}
\end{figure}

That said, cosmologically selected simulated galaxies cannot provide exact representations of all properties of the MW, and our simulations are no exception.
For example, these simulations span a range of two in stellar mass, and m12m, in particular, has stellar mass of $1.0 \times 10^{11} \Msun$, about twice the MW's stellar mass of $5 \times 10^{10} \Msun$ \citep{bhg16}, and closer to M31's stellar mass.
All three of these galaxies have higher cold gas masses and SFRs at $z = 0$ than the MW, though the MW does have lower SFR than typical disk galaxies of similar stellar mass at $z \approx 0$ \citep{Licquia2015}.
While the spatial disk structure of m12i, m12f, m12m are all similar to the MW, the kinematic structure is more analogous to M31 \citep{Dorman2015}, in particular, in the fiducial solar cylinder ($|z| < 0.3 \textrm{kpc}$, $7.95 < R/\textrm{kpc} < 8.45$, see section 3.2 for details), the magnitude of the total \emph{stellar} velocity dispersion is larger than the MW \citep[see][]{Nordstrom2004} at all ages (see Figure~\ref{fig:sigma}).
We explore the physical processes that set this velocity dispersion at birth and cause it to increase it with time in Loebman et al., in prep.
As with all such simulations, we are also ultimately limited by computational resources, and forced at some point to implement models for physical processes that occur below the resolution limit of the simulation. An extensive discussion of these models and the tests underlying the decisions made in this implementation are provided in \citet{Hopkins2017fire}.
Finally, because we compute dust extinction directly from the gas metallicity distribution in the simulations, the extinction map will differ from that of the MW.
Thus, we caution those using these simulations for mock MW catalogs to be aware of these differences.

\begin{deluxetable*}{lccccc}

\tablecaption{Halo-wide properties of our simulations at $z = 0$}
\tablehead{
 	Name &
	$N_{\rm particle}$ &
	$\Mthm \, \left[ M_\odot \right]$ &
	$\Rthm \, \left[ \kpc \right]$ &
	$R_{-2} \, \left[ \kpc \right]$ &
	$M_{\rm *,total} \, \left[ M_{\odot} \right]$
}

%\decimals

\startdata
\texttt{m12i} & 50,800,000 & 1.2e12 & 336 & 12.3 & 7.3e10 \\
\texttt{m12f} & 74,400,000 & 1.7e12 & 380 & 14.1 & 9.7e10 \\
\texttt{m12m} & 74,500,000 & 1.6e12 & 371 & 10.7 & 1.3e11 \\
\enddata

\tablecomments{
$N_{\rm particle}$: total number of dark matter, gas, and stars particles within $\Rthm$.
$\Mthm, \Rthm$: mass and radius that enclose 200 times the mean matter density.
$R_{-2}$: radius where log-slope of dark matter density profile is -2.
$M_{*,total}$: total stellar mass within $\Rthm$.
}
\label{tbl:halos}
\end{deluxetable*}

\begin{table*}
\begin{center}
\caption{Properties of the MW and Simulated Galaxies in This Paper.}
\begin{tabular}{lllllllc}
\hline
& $N_*$ & $M_* $ & $R_{*, 90}$ & $Z_{*, 90}$ & $R_{*,e}$\tablenotemark{b} & $M_{\rm gas}$ & SFR \\
Galaxy & & [$M_\odot$] & [kpc] & [kpc] & [kpc] & [$M_\odot$] & [$M_\odot \, {\rm yr}^{-1}$] \\
\hline
Milky Way\tablenotemark{a} & \nodata & $5 \pm 1e10$ & \nodata & \nodata & $2.6 \pm 0.5$ & 0.7e10 & 1.7 \\
\texttt{m12i} & 9,000,000 & 5.5e10 & 8.6 & 2.1 & 2.7 & 0.8e10 & 3.5 \\
\texttt{m12f} & 11,000,000 & 6.9e10 & 11.9 & 2.1 & 3.4 & 1.2e10 & 4.8 \\
\texttt{m12m} & 15,800,000 & 1.0e11 & 11.6 & 2.3 & 3.2 & 1.5e10 & 7.0 \\
\hline
\end{tabular}
\end{center}
\tablecomments{
$N_{*}$: number of star particles in the galaxy\tablenotemark{c}.
$M_{*}$: stellar mass within the galaxy.
$R_{*, 90}$: radius that encloses 90\% of stellar mass.
$Z_{*, 90}$: vertical height that encloses 90\% of stellar mass.
$M_{\rm gas}$: mass of gas within the galaxy.
SFR: star-formation rate within the galaxy, averaged over the last 100 Myr.
The simulated galaxies have higher SFRs than the MW because of both higher gas mass and higher gas surface density (Table~\ref{tbl:solar-circle}).
}
\tablenotetext{a}{Values from \citet{bhg16}. We could not find a value in the literature for $R_{*, 90}$ or $Z_{*, 90}$ in the MW.}
\tablenotetext{b}{For the MW, this is the scale radius for the ``thin'' disk.
For the simulations, we determine the scale radius via an exponential fit using star particles at $|Z| < 300$ pc of the disk plane (see \S \ref{subsec:galactocentric}) and $6 < R < 12$ kpc (thus excluding the bulge contribution), though the exact radial range of the fit does not significantly affect the result.}
\tablenotetext{c}{We define the extent of the galaxy (disk) by iteratively and simultaneously solving for $R_{*, 90}$ and $Z_{*, 90}$ using all star particles within 20 kpc. Convergence takes 10--20 iterations.}

\label{tbl:galaxies}
\end{table*}

\section{Coordinate systems}
\label{sec:coords}

Within each simulation snapshot, we establish a galactocentric coordinate system and choose a ``solar viewpoint;'' specifically, a phase-space position for the local standard of rest (LSR) of each synthetic survey.
First we determine the ``galactocentric'' coordinate frame, centered on and aligned with the stars in the galaxy. We represent galactocentric coordinates using lowercase $\vec{x}$, $\vec{v}$. Then we assume a phase-space location for the LSR in the simulation, to establish a system of ``LSR'' coordinates that we represent with capital $\vec{X}$, $\vec{V}$ notation in Cartesian or cylindrical coordinates or equivalently using angular coordinates $\ell, b$ for longitude and latitude respectively and $D$ for LSR-centric distance.
In the real Milky Way, the Sun has a small positional offset from the exact midplane of the Galactic disk and a small velocity offset relative to the mean motion of young stars on circular orbits in the vicinity of the Sun (how the LSR is usually defined). The resolution of the simulations is not fine enough to distinguish a difference between the LSR and a solar position/velocity, so heliocentric and LSR coordinates are equivalent for the purposes of these synthetic surveys. In this section we describe the details of these transformations.

\subsection{Galactocentric coordinates}
\label{subsec:galactocentric}

We first determine the center position of each galaxy, using an iterative ``shrinking spheres'' method, recursively computing the center of mass of star particles in a sphere, reducing the radius by 50\% and re-centering on the new center of mass at each iteration.
We then measure the center-of-mass velocity of the galaxy using all star particles within 15 kpc of this center.
We define this location as $(\vec{x}, \vec{v}) = (\vec{0}, \vec{0})$.

We then rotate the centered simulation into the principal-axis frame of the galactic disk.
First we assume that the sun is located at $R_{\odot} = 8.2 \kpc$ \citep{bhg16} in all three simulations (see discussion in \S\ref{subsec:heliocentric}).
Then, to define the principal axes, we compute the moment of inertia tensor using young star particles (age $< 1 \Gyr$) inside of this radius.
This establishes the $Z$ direction perpendicular to the plane of the galactic disk. As is the case for standard Galactocentric coordinates, we choose the orientation of the $Z$ axis such that the total angular momentum of the galactic disk points in the $-Z$ direction, so that the simulated galaxy rotates clockwise. If the LSR position is then located on the $-X$ axis, the rotation of the galaxy carries it in the $+Y$ direction, consistent with the standard assumptions used for Galactic coordinate systems.

We choose to define the orientation of the disk plane using young stars rather than gas for two reasons.
First, the gas disks in the simulated galaxies can be more misaligned (or warped) than the young stars with the axis defined by all stars.
Specifically, we find that the principal axis defined via all stars versus via gas ($< R_{\odot}$) leads to typical differences of $0.5$, $1.1$, and $1.3$ degrees in the orientations of the disks of m12m, m12f, and m12i.
This effect is less pronounced in the MW, and we do not wish this misalignment to complicate interpretations of mock surveys of stars.
Second, we anticipate that \Gaia itself may permit independent measurements of the disk centerline using stars alone, a case that potentially could be tested using our mock surveys.
The principal axis defined using only young stars differs far less relative to the gas than using all stars together, less than 0.5 degrees in all three cases.
Thus, we use young stars as a compromise between using all stars and gas.

\subsection{Local standard of rest}
\label{subsec:heliocentric}

\begin{figure*}
\centering
\includegraphics[width=\textwidth]{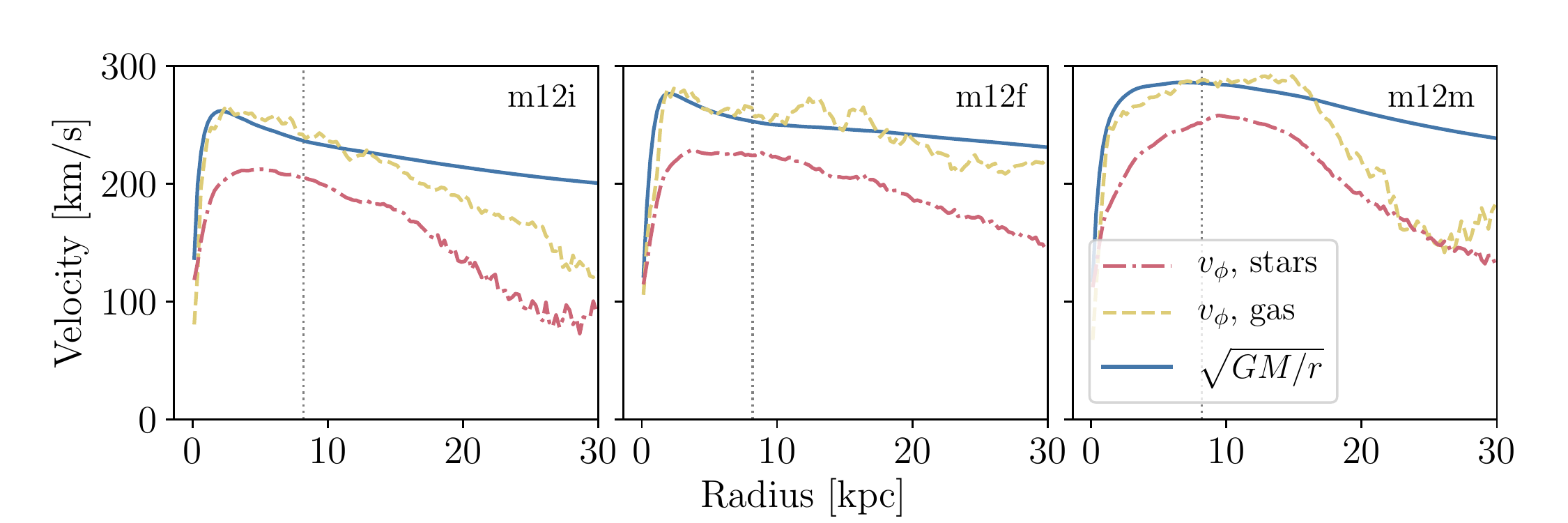}
\caption{Rotation curves ($\sqrt{G\sub{M}{tot}(<r)/r}$, solid blue) and bulk rotation velocity ($v_{\phi}$) curves for stars (red dot-dashed) and gas (yellow dashed) in the three simulations used to generate synthetic surveys. The vertical dotted black lines mark the solar circle, 8.2 kpc, chosen for the mocks. The MW's rotation curve peaks at 240 km \unit{s}{-1} \citep{bhg16}, which is within 10, 20, and 40 km \unit{s}{-1} of the peak value in the three simulations shown (from left to right). }
\label{fig:rotcurve}
\end{figure*}

Once the galaxy is centered and aligned, it remains to establish a coordinate system centered on a solar viewpoint and ``local standard of rest'' (LSR).
We first determine a suitable position for the ``solar circle'' $R_{\odot}$.
We explored whether to scale $R_{\odot}$ differently for each simulation based on disk scale radii, local density, or the local circular velocity, but there was no strong motivation for any of these possibilities, which would introduce significant extra complexity by requiring users to accommodate a different solar circle radius to study each different simulation. In all three simulations, using a multiple of the disk scale radius to locate the solar circle was within about 1.5$\sigma$ of the measured value in the MW, and there was more significant variation in choosing different azimuth locations (Loebman et al. in prep) than in choosing different radii within this range. The rotation curve ($\sqrt{GM(<r)/r}$) at 8.2 kpc is also flat and roughly consistent with MW values in each of the three simulations (Figure \ref{fig:rotcurve}). Thus we chose to prioritize simplicity and fix $R_{\odot}$ at the consensus value of $8.2 \kpc$ \citep{bhg16}. 
As a result, the azimuthally averaged local density in the ``solar neighborhood'' varies among the different simulations by roughly a factor of 3, which should be borne in mind when making comparisons with the MW.
Table \ref{tbl:solar-circle} summarizes relevant characteristics of the three simulations at $R_{\odot} = 8.2 \kpc$.

\begin{table*}
\begin{center}
\begin{tabular}{l|ll|llll|lll}
\hline
& \multicolumn{2}{c|}{surface density [$M_{\odot} \, \unit{pc}{-2}$]} & \multicolumn{4}{c|}{volume density\tablenotemark{d} [$10^{-3}\ M_{\odot}\ \, \unit{pc}{-3}$]} & \multicolumn3c{scale height [$\pc$]} \\
Galaxy &
total\tablenotemark{b} & baryonic\tablenotemark{c} &
total & stars & gas & DM &
stars & stars & cold\tablenotemark{e} \\
& & & & & & & thin & thick & gas \\
\hline
MW\tablenotemark{a} & 
$70 \pm 5$ & 
$47 \pm 3$ & 
$97 \pm 13$ &
$43 \pm 4$ &
$41 \pm 4$ &
$13 \pm 3$ &
$300 \pm 60$\tablenotemark{f} &
$900 \pm 180 $\tablenotemark{f} &
$150 $\tablenotemark{g} \\

\texttt{m12i} & 61 & 57 & 31 & 14 & 7.0 & 9.5 & 480 & 2000 & 800\tablenotemark{h} \\
\texttt{m12f} & 80 & 76 & 49 & 24 & 13 & 12 & 440 & 1280 & 360 \\
\texttt{m12m} & 145 & 152 & 100 & 54 & 31 & 17 & 290 & 1030 & 250 \\
\hline

\end{tabular}
\end{center}

\caption{
Azimuthally averaged properties at the ``solar circle,'' $R_{\odot} = 8.2$ kpc, for the three simulated galaxies, compared to properties of the Milky Way measured for the Solar neighborhood.
}
\label{tbl:solar-circle}
\tablenotetext{a}{Values from \citet{bhg16} except where otherwise noted.}
\tablenotetext{b}{Material within $|Z|\leq 1.1$ kpc (as in \citealt{bhg16}).}
\tablenotetext{c}{Material within $|Z|\leq 4$ kpc ($\sim$10 scale heights, to approximate $|Z| \to \infty$).}
\tablenotetext{d}{Material within $|Z|\leq 200$ pc (definition of ``local'' based on balance of volume \& sampling noise; see \S \ref{subsec:heliocentric}).}
\tablenotetext{e}{$T < 100$ K}
\tablenotetext{f}{\citet{Juric2008}}
\tablenotetext{g}{\textsc{Hi} gas, \citet{2008A&A...487..951K}}
\tablenotetext{h}{The azimuthally averaged gas vertical density profile in m12i is nearly constant to this height, though individual regions show smaller scale heights and dense clouds.}
\end{table*}

With $R_{\odot}$ fixed, we choose solar viewpoints at three evenly spaced azimuthal angles $\phi_{\odot}$ around each simulated galaxy, using a prime number of viewpoints to avoid selecting multiple viewpoints at the same relative position to features having azimuthal symmetries like bars ($m = 2$) or spiral arms (usually $m = 2n$). We place each viewpoint directly on the disk centerline ($Z_{\odot} = 0$), because the current consensus value of 10 pc for the Sun's height above the disk plane \citep{bhg16} is comparable to the force-softening length for star and gas particles in the simulation. 
To compute the LSR velocity, we find all star particles within 200 pc of each solar viewpoint.
We use 200 pc as a compromise between identifying a sufficiently ``local'' LSR and having enough star particles to avoid sampling noise.
We find that using 200 pc typically leads to $\sim 100$ star particles.
We then compute the median velocity using all star particles within 200 pc of each solar viewpoint, and we use this to define the LSR velocity vector.
Ideally we would use only young stars to set the LSR as well as the disk plane; however, we use all star particles to keep the estimate  as local as possible given the resolution of the simulation. In practice, the median age of the star particles used to estimate the LSR ranges between 1.7 and 3.6 Gyr, compared to a median age of $\sim$6 Gyr in the full simulated galaxies, so they are in fact reasonably young. In reality the Sun has a velocity of $\sim 10$ km \unit{s}{-1} in each Cartesian direction relative to the LSR \citep{bhg16}, but we place our solar viewpoints at identically the LSR velocity; this is a reflection of our inability to resolve the phase-space evolution of individual stellar populations below the resolution of our simulations.

Table \ref{tbl:solar-positions} lists the galactocentric phase-space coordinates for each solar viewpoint used to generate a mock catalog.
The mock catalogs contain positions in the LSR frame using Cartesian coordinates; to recover the galactocentric Cartesian coordinates use
\begin{eqnarray}
\label{eq:gc1}
\vec{x} &=& \vec{X} + \vec{x}_{\odot}, \\
\vec{v} &=& \vec{V} + \vec{v}_{\odot};
\label{eq:gc2}
\end{eqnarray}
that is, \emph{add} the appropriate vectors in Table~\ref{tbl:solar-positions} to the positions and velocities in the corresponding mock catalog.

\begin{table*}
\begin{center}
\begin{tabular}{l|ccc|ccc|ccc}
\hline
label & $x_{\rm LSR}$ & $y_{\rm LSR}$ & $z_{\rm LSR}$ & 
$v_{\rm x, LSR}$ & $v_{\rm y, LSR}$ & $v_{\rm z, LSR}$ &
$v_{\rm R, LSR}$ & $v_{\rm Z, LSR}$ & $v_{\rm \phi, LSR}$ \\
\hline
\hline
\decimals
\texttt{m12i-lsr0} & 0.0 & 8.2 & 0.0 & 224.7092 & -20.3801 & 3.8954 & -17.8 & -3.9 & 224.4 \\
\texttt{m12i-lsr1} & -7.1014 & -4.1 & 0.0 & -80.4269 & 191.7240 & 1.5039 & -24.4 & -1.5 & 210.9 \\
\texttt{m12i-lsr2} & 7.1014 & -4.1 & 0.0 & -87.2735 & -186.8567 & -9.4608 & 22.1 & 9.5 & 206.5 \\
\hline
\texttt{m12f-lsr0} & 0.0 & 8.2 & 0.0 & 226.1849 & 14.3773 & -4.8906 & 14.9 & 4.9 & 227.9 \\
\texttt{m12f-lsr1} & -7.1014 & -4.1 & 0.0 & -114.0351 & 208.7267 & 5.0635 & -3.4 & -5.1 & 244.3 \\
\texttt{m12f-lsr2} & 7.1014 & -4.1 & 0.0 & -118.1430 & -187.7631 & -3.8905 & -11.4 & 3.9 & 227.4 \\
\hline
\texttt{m12m-lsr0} & 0.0 & 8.2 & 0.0 & 254.9187 & 16.7901 & 1.9648 & 16.2 & -2.0 & 254.7 \\
\texttt{m12m-lsr1} & -7.1014 & -4.1 & 0.0 & -128.2480 & 221.1489 & 5.8506 & 2.4 & -5.9 & 252.7 \\
\texttt{m12m-lsr2} & 7.1014 & -4.1 & 0.0 & -106.6203 & -232.2056 & -6.4185 & 15.4 & 6.4 & 265.3 \\
\hline
\end{tabular}
\end{center}
\caption{
Phase-space coordinates of the local standard of rest (LSR), corresponding to each solar viewpoint, for the nine mock catalogs, in the galactocentric reference frame described in \S\ref{subsec:galactocentric}.
Positions are in kpc and velocities in km \unit{s}{-1}.
We also list velocities in cylindrical coordinates for reference.
See \S\ref{subsec:heliocentric} for discussion.
}
\label{tbl:solar-positions}
\end{table*}

\subsection{Elemental abundances}
\label{sec:abundances}

We report the simulation's 11 elemental abundances (H, He, C, N, O, Ne, Mg, Si, S, Ca, Fe) alongside the phase-space data for stars in the mock catalogs.
Because the simulations track the total mass of each element produced per particle (inherited by star particles as they form from enriched gas particles), we assume Solar values to convert to metallicities of the form $[\Xi/H]$.
We use the Solar values from \citet{asplund09} to calculate
\begin{equation}
[\Xi/H] \equiv \frac{m_\Xi/m_{\Xi,\odot}}{m_H/m_{H,\odot}}
\end{equation}
for each star particle and each element $\Xi$, where $m_{\Xi}$ is the mass of a given element associated with the star particle and $m_{\Xi,\odot}$ is its Solar value.

\section{Mock catalogs}
\label{sec:mock-catalog}

Cosmological simulations such as ours produce distributions of star \textit{particles}, each of which represents the position, velocity, and properties like age and metallicity for an IMF-averaged ensemble\footnote{Incorporation of stochastic IMF sampling, necessary only at much higher resolution than the simulations here, is ongoing; see Wheeler et al. in prep.} of stars. On the other hand, observed stellar properties from a survey like \Gaia are frequently a function of apparent magnitude and stellar type; i.e., the position of a given star on the Hertzsprung-Russell diagram and its LSR-centric distance. In order to create a true synthetic survey, therefore, one must follow an algorithm for creating a so-called ``mock catalog'' by generating a set of \textit{synthetic stars} from each star particle in the simulation.

In the case of the simulations here, star particles have an initial mass of $7070 \, M_\odot$ (and a typical mass of $\approx 5000 \, M_\odot$ at $z = 0$).
We therefore can treat the properties of each star particle as representing a population of stars, as is done in the simulation itself to follow stellar evolution and calculate feedback and metal enrichment. Specifically, we consider stars in the population represented by one star particle to have a single age and identical abundances, thus described by one model isochrone. 

To create a mock catalog, synthetic stars are spawned from each star particle in the simulated galaxy following the resampling algorithm outlined in \citep{sharma11}, which we adapt here for our purpose. In brief, we follow these steps:

\begin{enumerate}
\item Draw stellar masses adding up to the total mass of the star particle by sampling the \citet{Kroupa2001} initial mass function (IMF), consistent with the IMF used in the simulation (\S \ref{sec:imf}).

\item Compute the stellar properties and absolute Gaia $G$, $G_{Bp}$, and $G_{Rp}$ magnitudes of each synthetic star by interpolating in initial stellar mass over the isochrone with the closest metallicity and age from a model grid (\S \ref{sec:isochrones}). All stars produced from a given star particle are assigned the same age and elemental abundances as the generating particle, consistent with our interpretation of star particles as tracking single stellar populations. Stars with estimated apparent magnitudes in the \Gaia range ($3<G<21$, ignoring extinction) are added to the mock catalog.

\item Place each synthetic star in phase space by sampling from a locally varying one-dimensional kernel in each of position and velocity space, centered on the generating star particle. To achieve greater dynamic range in phase-space density, we use a series of different density kernels for stars of different ages (see \S \ref{sec:kernel-sampling}).
\end{enumerate}

The resulting mock catalog contains all stars that would fall in the \Gaia survey if there were no Galactic extinction. Below we describe the specific assumptions used to generate the stellar parameters (using an IMF and model isochrones) and phase-space positions (using a density estimator) of the synthetic stars.

\subsection{Initial mass function}
\label{sec:imf}

We assume that the initial masses $m_i$ of the synthetic stars represented by each star particle are distributed according to the initial mass function (IMF) represented by the continuous PDF
\begin{equation}
p(m_i) dm_i = \mathcal{N} \left\{ \begin{array}{ll}
m_i^{-0.3} & 0.01<m_i<0.08 M_{\odot} \\
0.08 m_i^{-1.3} & 0.08<m_i<0.5 M_{\odot} \\
0.04 m_i^{-2.3} & 0.5<m_i
\end{array} \right. ,
\end{equation}
as proposed in \citet{Kroupa2001}, where $\mathcal{N}$ is the normalization factor such that 
\begin{equation}
\int_{0.01 M_{\odot}}^{120 M_{\odot}} p(m_i) dm_i = 1.
\end{equation}
In practice, although the IMF is defined over the full range $0.01<m_i<120 M_{\odot}$, we draw stars from a subrange of this IMF. For efficiency, we define the minimum mass $m_i^{\mathrm{min}}$ associated with each star particle individually, by setting the minimum absolute magnitude corresponding to the faint apparent magnitude limit of the survey at the distance of the star particle, allowing for the local size of the kernel used to spread out the synthetic stars. 

In the simulation the star particle's mass represents the total mass of stars \emph{currently} part of the associated stellar population, excluding mass lost to winds and stars that have become stellar remnants. This sets an upper limit $m_i^{\mathrm{max}}$ on the range of the IMF to be sampled based on the prediction of the isochrone with the appropriate age and metallicity for the most massive star that is still part of the population. In practice we use the closest isochrone in the model grid to set $m_i^{\mathrm{max}}$. This introduces a slight inconsistency in the sense that the isochrones used to create the mock catalogs (from the Padova group, \citealt{marigo17}) are not the same set as were used in the simulation, which uses STARBURST99 v7.0 to track stellar evolution and mass loss via the Geneva evolutionary tracks for high-mass stars \citep{Leitherer1999,leitherer14}. These two sets of isochrone models take into account different subsets of effects that control the evolution of high-mass stars: for example the version of STARBURST99 used in the simulation includes enhanced mass loss from rotation, while the \citet{marigo17} isochrones include thermal pulsations from AGB stars. However, the effect of this discrepancy is probably mitigated by our assumption that the mass-loss rate is mass-independent, which we use to compute the total number of stars to sample from each particle. Nevertheless users wishing to do detailed studies of the synthetic evolved stellar populations should keep this limitation in mind.

To determine the number of stars to sample from each star particle, we compute the fraction of the IMF within the subrange being sampled,
\begin{equation}
\sub{f}{IMF} = \int_{m_i^{\mathrm{min}}}^{m_i^{\mathrm{max}}} p(m_i) dm_i,
\end{equation}
and estimate the number of stars associated with the star particle using the median initial stellar mass $\mathfrak{m}_i$, defined by
\begin{equation}
\frac{1}{2} = \int_{0.01 M_{\odot}}^{\mathfrak{m}_i} p(m_i) dm_i.
\end{equation}
Assuming a mass-loss rate independent of stellar mass, this implies that the \emph{total} number of stars to be sampled over the entire IMF for a star particle with mass $m_p$ is
\begin{equation}
\sub{\tilde{N}}{tot} = \frac{m_p}{\mathfrak{m}_i}
\end{equation}
and the number in the subrange is
\begin{equation}
\tilde{N}  = \sub{f}{IMF} \frac{m_p}{\mathfrak{m}_i}.
\end{equation}
Because $\tilde{N}$ is not an integer we use stochastic rounding to determine the integer number $N$ of stars to generate by drawing a random number $u$ from a uniform distribution on $[0,1)$ and setting:
\begin{equation}
N = \left\{ \begin{array}{ll}
\lfloor \tilde{N} \rfloor+1 & \tilde{N} - \lfloor \tilde{N} \rfloor \geq u \\
\lfloor \tilde{N} \rfloor & \textrm{otherwise}
\end{array}\right.
\end{equation}
Because our particle mass, $m_p \sim 7000 M_{\odot}$, is nearly two orders of magnitude larger than the maximum allowed stellar mass on the IMF (120 $M_{\odot}$), we consider this approach adequate to represent the high-mass end of the IMF.

The initial mass of each of $N$ synthetic stars belonging to the generating particle is then sampled from the IMF between $m_i^{\mathrm{min}}$ and $m_i^{\mathrm{max}}$.
We use this mass to place the synthetic star on a model isochrone, which determines its magnitude, color, and stellar parameters.

\subsection{Isochrones}
\label{sec:isochrones}

To describe stellar populations we use the PARSEC model isochrones (release v1.2S and COLIBRI release PR16), as in \citet{marigo17} (see also \citealt{bressan12} and \citealt{marigo13}). We use isochrone tables computed for the \Gaia DR2 photometric system, including models for O- and C-rich circumstellar dust\footnote{Specifically, the \texttt{dpmod60alox40} model for O-rich dust and the \texttt{AMCSIC15} model for C-rich dust} from \citet{groenewegen06}. These were tabulated with CMD 3.0\footnote{\url{stev.oapd.inaf.it/cmd}} for a grid of 34 metallicities from $Z = 0.0001$ (1/300th solar) to $Z = 0.03$ (roughly twice solar) and 71 age values evenly spaced from $\log_{10}$(age/yr)$= 6.6$ (3.98 Myr) to $\log_{10}$(age/yr)$= 10.1$ (12.6 Gyr). These isochrone tables extend to a minimum stellar mass of $0.08\,M_{\odot}$; they do not include a model of the instability strip or the white dwarf sequence.

\begin{figure*}
\begin{center}
\plottwo{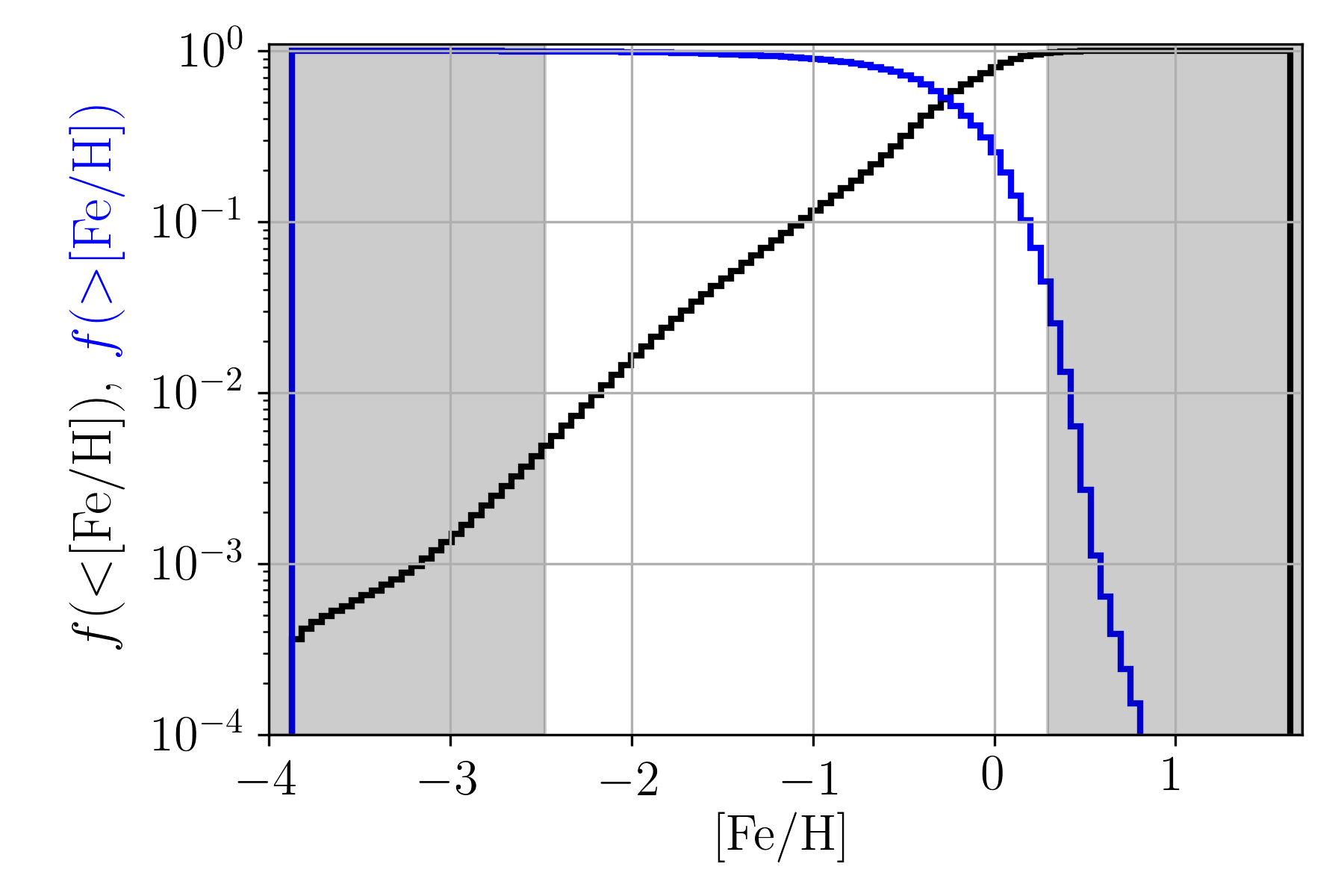}{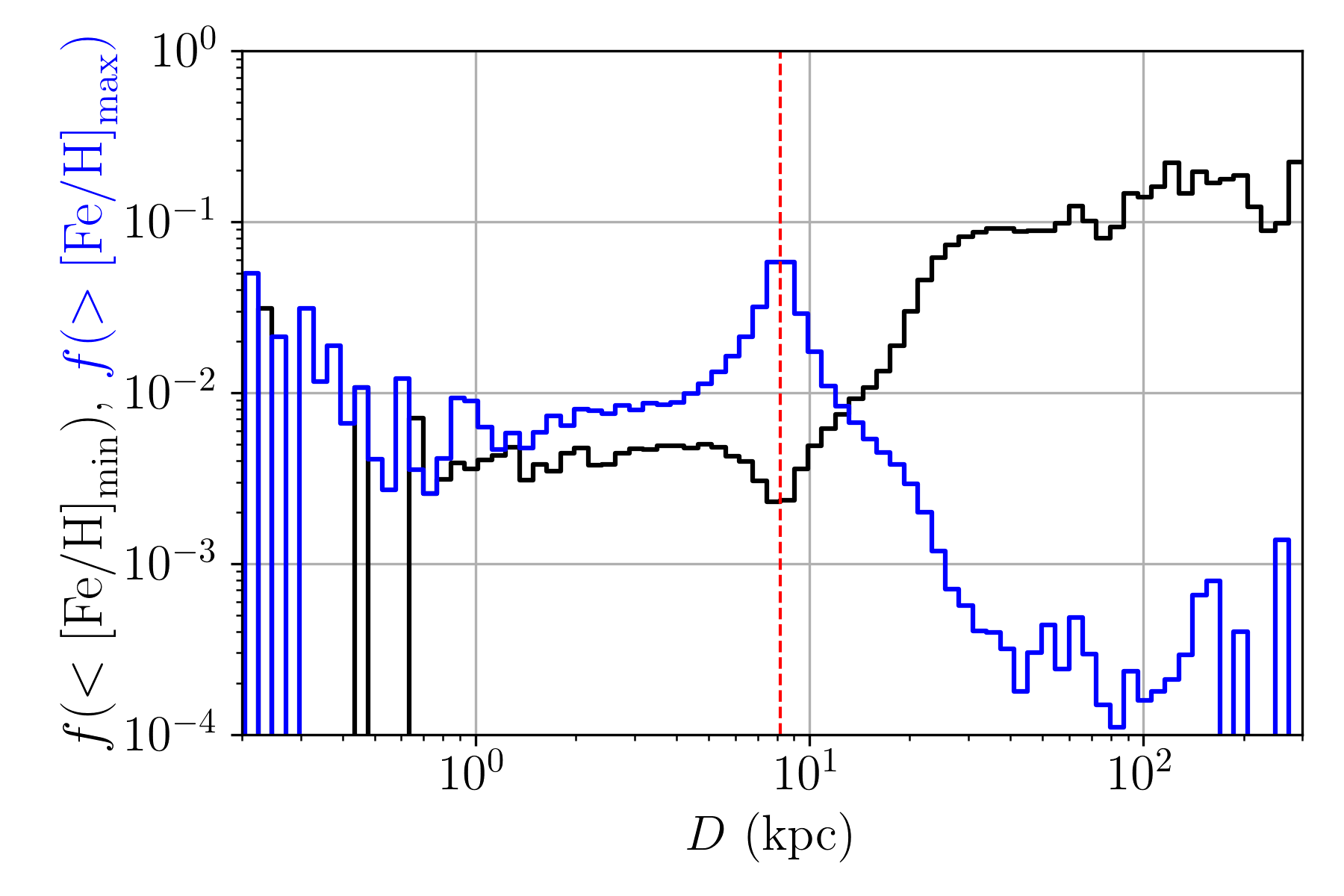}
\caption{Most star particles in the simulated galaxies have [Fe/H] within the isochrone model grid used to generate the mock catalogs.
\textbf{Left:} Cumulative (black) and reverse-cumulative (blue) distributions of [Fe/H] for star particles in \texttt{m12i-0} within 300 kpc of the solar position (established as described in \S \ref{sec:coords}). The model isochrone series used to create synthetic stars spans the [Fe/H] range shown in white; simulated star particles in the grey-shaded regions are assigned to the isochrone at the edge of the grid.
\textbf{Right:} Fraction of star particles falling above (blue) or below (black) the [Fe/H] range of the model isochrones, as a function of distance. The regions containing the most outliers are either in the bulge (red dashed line), which in the final synthetic survey is heavily extincted, or in the distant halo, where few stars will be bright enough to enter the synthetic survey.}
\label{fig:metallicity-distribution}
\end{center}
\end{figure*}

Simulated star particles have abundances that range from the initial floor of [Fe/H]$= -4$ to a maximum of [Fe/H]$= 1.5$; only a few percent of these fall outside of the range spanned by the model isochrones (Figure~\ref{fig:metallicity-distribution}, left panel).
For these we use the isochrone at the appropriate edge of the model grid.
Extremely low [Fe/H] values below the edge of the model grid are likely seriously affected by stochastic noise and the choice of the floor value, so we counsel users to treat synthetic stars at these abundances with caution. High-[Fe/H] outliers mainly occur in the bulge, which is significantly extincted in the synthetic survey; low-[Fe/H] outliers are only a significant contributor in dwarf satellite galaxies, which will have relatively few stars bright enough to enter the survey volume (Figure \ref{fig:metallicity-distribution}, right panel). Thus we consider this model grid to be sufficiently broad for most users of the synthetic survey.

\subsection{Phase-space density}
\label{sec:kernel-sampling}

We spread out the synthetic stars in six-dimensional phase space relative to their generating star particle using a kernel $\mathcal{K}(r)$ with the parabolic or \citet{epan69} form
\begin{equation}
\mathcal{K}(r) dr \propto (1-r^2) r^5 dr,
\end{equation}
normalized to integrate to 1 over the kernel volume in six dimensions. The kernel radius $r$ is related to the smoothing length in the $i$th dimension $h_i$ by 
\begin{equation}
r=\sqrt{\sum_{i=1}^{6} (\delta y_i/h_i)^2},
\end{equation}
where ${\bf \delta y}$ is the distance vector in 
the six dimensional phase space relative to the generating star particle. 
The phase space is defined in the Cartesian position and velocity dimensions of the LSR frame, $(X,Y,Z,V_X,V_Y,V_Z)$, and the kernel size is computed in this coordinate system at the point of each star particle using the EnBiD scheme \citep{2006AJ....132.1645S} such that the local density in position and velocity space varies smoothly from particle to particle \citep[see][for a full discussion]{sharma11}.
The scheme computes a locally adaptive metric making use of a binary space partitioning tree scheme, where the partitioning criterion is determined by comparing the Shannon entropy or information along different dimensions.
The scheme was further refined in the EnLink code for the purpose of clustering analysis \citep{sharmaEnlink}.
We here use the EnLink code (instead of the publicly available EnBiD code) and employ the cubic cell scheme of EnBiD, which means that $h_1=h_2=h_3$  and $h_4=h_5=h_6$. For the two independent smoothing lengths we use the geometric mean of the smoothing lengths along each of the three dimensions. Relative to a fully multivariate six-dimensional phase-space kernel, this approach allows for faster and less noisy sampling. For this work we use the nearest eight neighboring star particles to compute the kernel size.

\begin{figure*}
\begin{center}
\includegraphics[height=0.8\textheight]{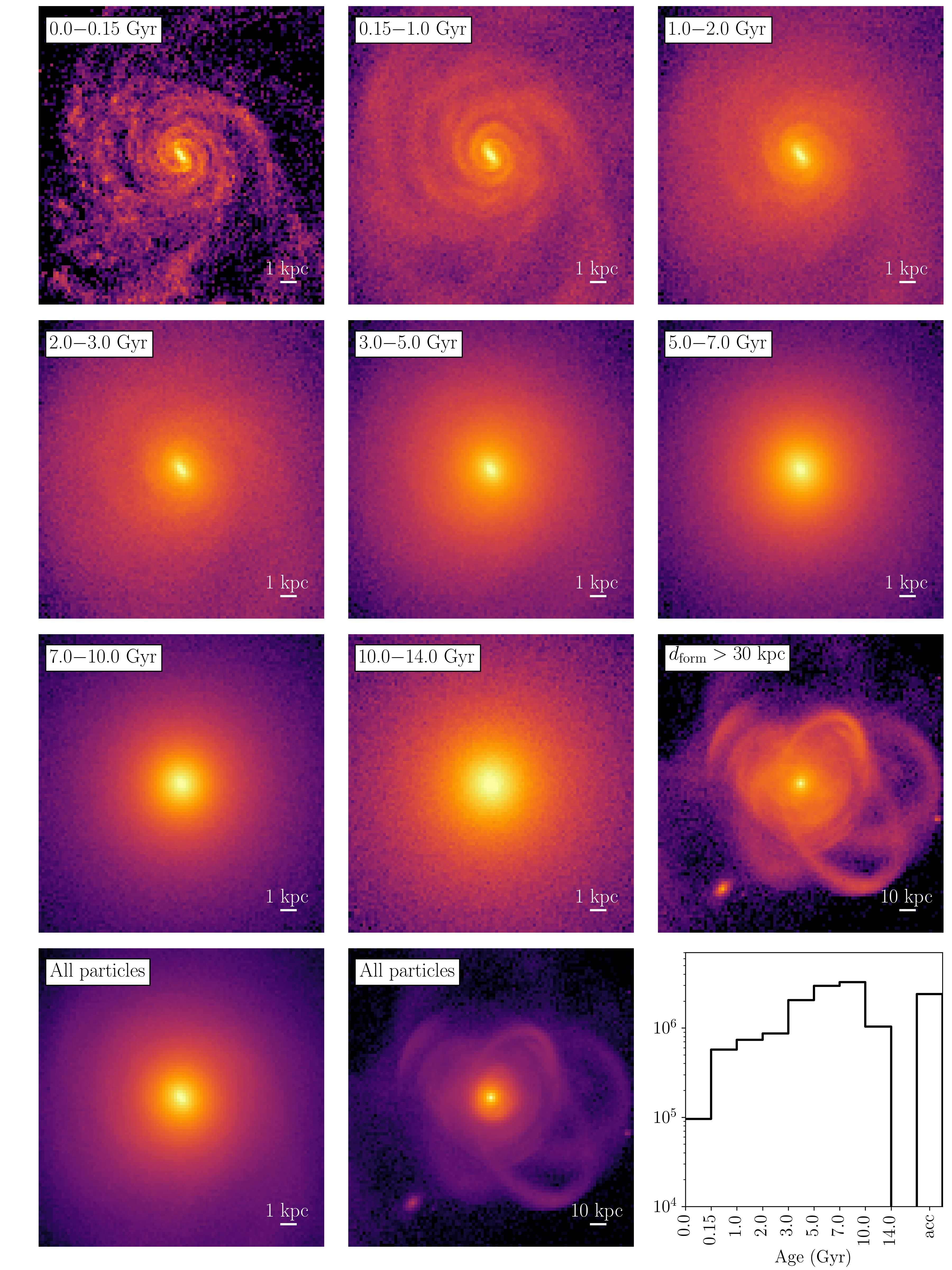}
\end{center}
\caption{Different star particle subsets used to construct density kernels for resampling the phase-space distribution of \texttt{m12i}. The top three rows show the projected density in the different age bins of in situ stars (those formed within 30 kpc of the main halo) and accreted stars (formed beyond 30 kpc from the main halo). Each panel's color range is independently adjusted to show the full range of projected densities in the subset. The bottom left and center panels compare the total density distribution if all the particles are considered together (left) versus the superposition of the different age-dependent kernels shown in the top three rows (center).  Substructure apparent in many of the subsets is mostly washed out if the entire distribution is used to create one kernel. The bottom right panel shows that each bin in the top three rows still contains at least $10^5$ particles, and usually more than $10^6$, to sample the phase-space density.}
\label{fig:subsampling}
\end{figure*}

In general, an $N$-body system is composed of various stellar populations each having its own unique phase-space density. If the phase-space density is calculated by treating all the $N$-body particles as a single population, fine phase-space structures in the distribution that overlap with interlopers from kinematically hotter populations in either position or velocity space will be oversmoothed \citep[for further details see][]{sharma11}. To overcome this, in \citet{sharma11}, while sampling the $N$-body stellar halo of \citet{bj05}, each satellite galaxy was treated as a separate population.
In our case the phase-space properties change with age. Hence, to mitigate this problem of oversmoothing, we generate separate kernel maps for stars formed in situ (within 30 physical kpc of the main galaxy; see \citealt{Bonaca2017} and \citealt{Sanderson2017c}) in the eight age bins corresponding to the populations of the Besan\c con Milky Way model \citep{Robin2003}, and a separate kernel for all stars formed further than 30 kpc from the main galaxy. Star particles belonging to each of these subsets are resampled to produce synthetic stars using the kernel generated from that subset alone. This strategy allows us to better resolve cold phase-space structures, especially in a few key contexts: young stars in the thin disk, triaxial features like bars in the galactic center, and tidal streams in the halo. Figure~\ref{fig:subsampling} illustrates the advantage of this density resampling strategy using \texttt{m12i}. The first three rows show the projected density distribution for each subset of particles used to generate a kernel. The dense, kinematically cold structures in the thin disk and halo are clearly seen, as is a triaxial component in the center of the galaxy. The hotter, older populations, which still change shape slightly with age, can also be faithfully represented. Many of these interesting features would vanish if all particles were considered together (bottom left and center panels). Our subdivisions are sufficiently few that each density distribution is well resolved (bottom right panel).

Our choice of phase-space smoothing that is adaptive with age mitigates one issue associated with creating mock catalogs from simulations, but cannot mitigate the fundamental limitation of the simulation resolution. Thus all stars sampled from each particle, assumed to be part of a single stellar population (SSP), necessarily represent an approximation to the phase-space distribution of that SSP, the evolution of which is not resolved. We thus caution users of the catalogs against over-interpreting detailed phase-space substructure in the catalogs below the scale of the SSP model sampling. In order to aid in identifying this limit we provide the index of the parent star particle for each star in the catalog: structures made entirely of stars with the same ``parent ID'' should be considered spurious.

\section{Synthetic survey}
\label{sec:synthetic-survey}

The mock catalog described in \S\ref{sec:mock-catalog} includes perfect information for all the synthetic stars that would be observed by \Gaia in the absence of galactic extinction. To create a true synthetic survey, we must include models for extinction and observational uncertainties, in order to determine which stars actually fall in the magnitude range accessible to \textit{Gaia}. These stars constitute what we call a ``synthetic \Gaia survey'' of one of the simulated galaxies, for a particular solar viewpoint. In this section we summarize the models used to determine extinction and observational uncertainties.

\subsection{Extinction modeling}
\label{subsec:extinction}

The simulations follow the evolution of gas and its metallicity in each simulated galaxy, from which stars are formed according to the criteria discussed in Section \ref{subsec:numerics}. 
Since we do not resolve the creation and destruction of dust grains, we infer the line-of-sight extinction by dust by assuming that it traces the metal-enriched gas. Our approach is in contrast to that of previous mock catalogs, which have imposed the empirical model of the MW itself to provide interpolated extinction and reddening. While this was appropriate for models like Besan\c con and TRILEGAL, which are intended to reproduce the stellar density and structure of the MW, it is not accurate for simulated galaxies. Indeed, one of the advantages of cosmological simulations featuring detailed models for stellar feedback and ISM physics (as in our FIRE-2 simulations) is the ability to capture features like massive young clusters and spiral arms in the disks of the simulated galaxies. Correctly incorporating these features into a synthetic survey requires properly estimating the extinction from the gas and dust that are (physically) highly correlated with them. Likewise it is well known that dust is specifically correlated with locations of star formation and massive young stars, both locally (in GMC-scale environments) and globally (in the scale-lengths and heights of the cold gas, dust, and star-forming disks). Artificially imposing the MW extinction map on stellar populations and star-forming structures that are not in the same locations as the MW can introduce a large number of un-physical features and potentially serious biases. We therefore base our extinction model on the {\em same} gas distribution that the simulations calculate and use to determine where stars form.  We refer to this for short as a ``self-consistent'' dust model in the sense that it uses information about the gas and metals distribution in the simulated galaxy, and thus captures the spatial correlations between extinction and star formation to the extent permitted by our simulations.

We compute the integrated column density of hydrogen, weighted by its total metallicity (from all species that we track),
\begin{equation}
N_H^{\rm eff} = \int_{0}^{r_*} (\sub{n}{H1,gas} + 2\,\sub{n}{H2,gas})\,\left(\frac{Z_{\rm gas}}{Z_{\odot}}\right) d\ell,
\end{equation}
along the line of sight $\ell$ from the solar position adopted for the mock catalog (defined as the origin using the transformations discussed in \S\ref{sec:coords}) to the location of the star particle, $r_*$. Here $\sub{n}{H1,gas}+2\,\sub{n}{H2,gas}$ is the total (atomic+molecular) number density of neutral hydrogen atoms in the gas, calculated in-code, weighted by $Z_{\rm gas} / Z_{\odot}$, the ratio of the local total metal mass to solar.
We assume a constant dust-to-metals ratio in the neutral gas, rather than a constant dust-to-gas ratio, to account for variations in the gas metallicity along the line of sight.
We compute this integration for all star particles within 500 kpc of the galactic center (as defined in \S\ref{subsec:galactocentric}).

Using this densely sampled, three-dimensional map of the column density, we calculate the column density to each synthetic star either by direct assignment (if the synthetic star is located at the exact position of its generating particle) or by linear interpolation with its closest neighbors in a three-dimensional Delaunay triangulation \citep{qhull,scipy}. We then calculate the corresponding $B$-$V$ reddening according to 
\begin{equation} \label{eq:reddening}
\frac{N_H^{\rm eff}}{E(B-V)} \equiv \sub{Q}{dust} \sim 2.5 \times 10^{22}\ \mathrm{H}\ \unit{cm}{-2}\ \unit{mag}{-1}.
\end{equation}
We intentionally choose a conservatively high value of \sub{Q}{dust}, such that it produces a \textit{smaller} amount of reddening than median observational estimates in the MW \citep[but within the range of systematic uncertainties and line-of-sight variance; see][who find $\sub{Q}{dust}\sim 0.3-5\times10^{22}$ in these units]{2011A&A...533A..16W,2012ApJS..199....8G,2013MNRAS.431..394W,2018arXiv180511787N}. We err on the side of possibly under-estimating the reddening and extinction for three reasons. First, this includes {\em more} stars in the synthetic survey, allowing users to increase the extinction or vary the input extinction curve, if desired, while maintaining completeness of the catalog (\S \ref{subsec:reprocessing} explains how to do this; see also Figures \ref{fig:extinction_age} and \ref{fig:extinction_map} and the discussion in \S \ref{sec:results}). Second, some authors have argued this value better represents many nearby MW-mass galaxies, to which our galaxies may be better proxies \citep[see][]{2018ApJ...855..133K}. 
%Third, this choice produces a distribution of $E(B-V)$ versus latitude in the simulations that better match the MW \citep{drimmel}, likely because our simulated galaxies are more gas-rich (with higher column density) than the MW.

We use the standard law
\begin{equation}
A_{0} = 3.1 E(B-V)
\end{equation}
to calculate the extinction. This relation is consistent with the assumption used in \citet{2018arXiv180409378G}, which gives the formulae for transformations from the global extinction to extinction in the \Gaia DR2 passbands that we then use to calculate $A_G$, $A_{G_{BP}}$, and $A_{G_{RP}}$. We use these extinctions to calculate the observed (extincted) magnitudes and reddened colors of synthetic stars, to determine which stars fall in the \Gaia apparent magnitude range. 

The un-extincted catalogs are all at least 2.5 times the size of the final synthetic surveys and therefore even more unwieldy to deal with, and due to our choice of a fairly high value of \sub{Q}{dust}, we expect that only a small number of users will need them. Therefore, in the interest of avoiding confusion between similar data sets, we have currently released publicly only the final synthetic surveys (which we emphasize $do$ include all the information required to recalculate different extinction and reddening coefficients).  However, we are happy to provide the complete un-extincted catalogs upon request.

\subsection{Error modeling and selection function}

The final step in producing a synthetic survey is to simulate the observational uncertainties using an error model. \Gaia errors have spatially complex structure because of the scanning law that determines the number of times each star transits its two telescopes, compounded by further choices in filtering these transits to determine the subset used to produce the final measurements. Additional complexity arises in crowded fields from, for example, deciding which stars participate in the full five-parameter astrometric solution. Therefore, the \Gaia error model continues to evolve from its pre-launch characterization, but no updated error model for DR2 is yet available. For our first release, we therefore choose simplicity in representing the observational uncertainties and selection function, while also reporting the underlying or ``true'' values of the observables for each star, so users can apply more sophisticated models of the \Gaia errors to the synthetic survey as needed. We hope to do the same in a future mock catalog release.

\subsubsection{Selection function}
\label{subsec:selection}

We include stars in the synthetic survey whose error-convolved, extincted apparent \Gaia $G$ magnitude is in the range $3 < G < 21$. The error model thus influences which stars end up in the final catalog in a limited way, especially given that the estimated photometric uncertainty at $G = 21$ for DR2 is approximately 10 mmag. This range is consistent at the faint end with the magnitude at which completeness drops below 90\% in DR1, and with the faint limit for stars that are consistently included in the five-parameter astrometric solution in DR2. We strongly discourage any over-interpretation of comparisons at the faint end of the synthetic surveys. At the bright end there are few stars in the synthetic surveys; we urge users of the mock catalogs in this regime to keep in mind the caution of the \Gaia team that the completeness is likely quite uneven for bright stars in the real \Gaia survey.

\Gaia measures radial velocities for a subset of bright stars. In DR2, and in our synthetic surveys, RVs are reported for stars with $G_{RVS} < 14$ and estimated effective temperatures in the range $3550 < \sub{T}{eff} < 6900$ K. 

The true \Gaia selection function is far more complex than this. We chose to use a simple selection (magnitude and, for RV data, temperature cuts alone) to avoid interference with more detailed selection function models, which are currently a work in progress; as they become available, we will apply to our synthetic surveys. \textit{We encourage users of the synthetic surveys to consider how their applications may be affected by the Gaia selection function}, and to apply the appropriate level of modeling for their science case.

\begin{table*}
\begin{center}
\begin{tabular}{p{3in}lll}
\hline
Quantity &  $a$ & $b$ & $c$\\
\hline
\hline
Photometric uncertainty in $G$ & 0.000214143 &  $1.07523\times 10^{-7}$ &  1.75147  \\
Photometric uncertainty in $G_{Bp}$, $G_{Rp}$ & 0.00162729 &  $2.52848\times 10^{-8}$ &  1.25981\\
Astrometric uncertainty in RA, dec, parallax & 0.0426028 &  $2.583\times 10^{-10}$ &  0.923162\\
Astrometric uncertainty in proper motion &  0.0861852 & $6.0771\times 10^{-9}$ &  1.05067 \\
Spectroscopic uncertainty in radial velocity &  0.278939 &  0.0000355589 &  1.10179\\
\hline
\end{tabular}
\end{center}
\caption{Coefficients of the simplistic Gaia DR2 error model used for the synthetic surveys.}
\label{tbl:errormodel}
\end{table*}

\subsubsection{Error model}
\label{subsec:errormodel}

The true error model for \Gaia DR2 is a complex function of sky position, magnitude, color, and temperature that involves the spacecraft scanning law and a full characterization of the reduction processes for the different astrometric, photometric, and spectroscopic measurements contributing to the final catalog. Preliminary versions of simplified models for the uncertainties have been provided in the public software package \textsf{pygaia}\footnote{\url{https://github.com/agabrown/PyGaia}}, but as this package has not been updated since DR2, the error prescriptions in this package differ significantly in some cases from the behavior noted in the DR2 release paper \citep{DR2}. We anticipate, moreover, that improved characterizations of the DR2 error model, as well as those for subsequent data releases, will become available in the future, and we wish to encourage the application of these to the mock catalogs.

Instead, we provide with this release a single draw from an error model described by a simple set of functions calibrated to the characterizations in \citet{DR2}. In all cases we model the uncertainties as solely dependent on apparent $G$ magnitude using an exponential form,
\begin{equation}
\sigma_i = a_i + b_i \exp(G/c_i),
\end{equation}
fit to the estimates given in the paper. The coefficients used for the uncertainties on different quantities are listed in Table \ref{tbl:errormodel}. For the radial velocities we add a systematic noise floor of 0.11 km \unit{s}{-1} as described in the documentation for the \Gaia DR2 data model.\footnote{\url{https://www.cosmos.esa.int/documents/29201/1645651/GDR2_DataModel_draft.pdf}}

For convenience, we provide both the underlying values and one set of ``error-convolved''  values (random draws from one-dimensional Gaussian distributions centered on the underlying values) for \Gaia observables as well as error estimates using our simple model. We caution that, given the simplistic nature of our error model, the error-convolved values should not be used for detailed explorations of, for example, the \Gaia selection function, or analysis near the magnitude limit of the survey. The error-convolved values are intended for convenience, to understand broadly how the magnitude-dependence of the observational errors affects results in an order-of-magnitude sense.

\begin{figure*}
\begin{center}
\includegraphics[width=\textwidth]{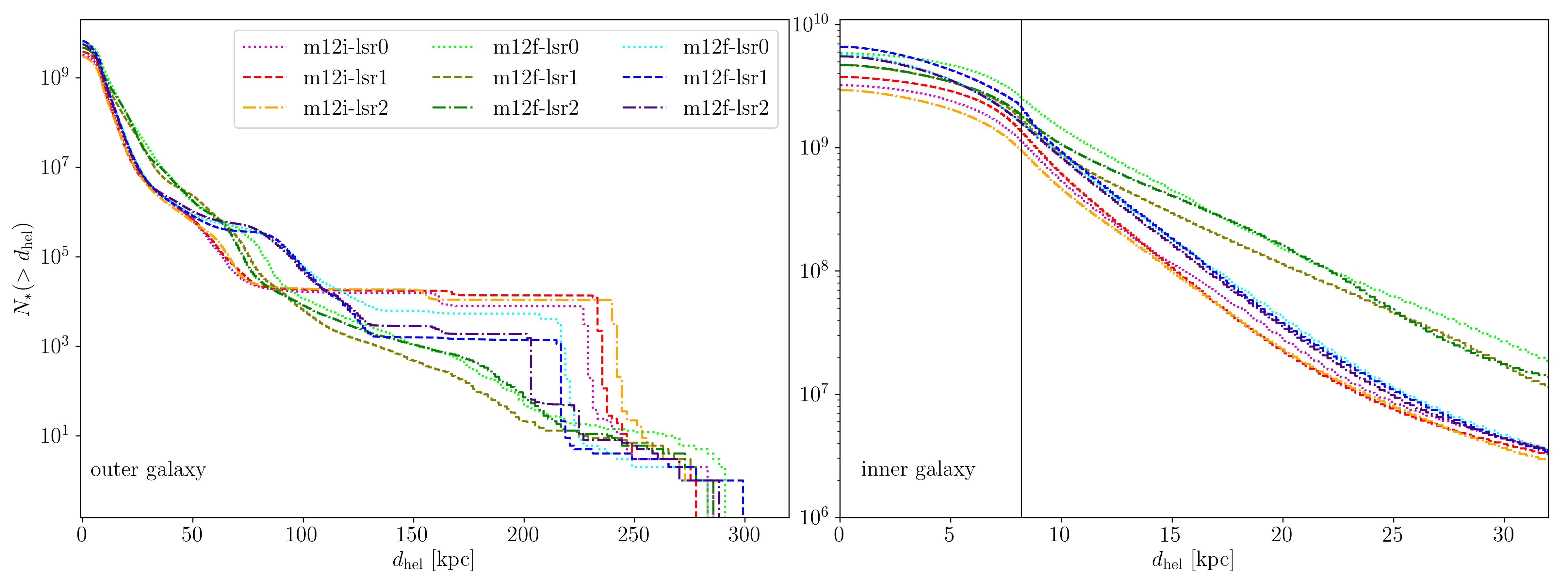}
\end{center}
\caption{The true distribution of stars in the synthetic surveys shows broad variety in both the inner and outer galaxy. \textbf{Left}: a view of the full extent of the synthetic surveys shows that they include at least a handful of stars at the edge of the simulated galaxy, and show significant variation because of accreted substructure. Steep drop-offs in the cumulative histogram correspond to bound satellite galaxies.
\textbf{Right}: A view of the inner galaxy shows the variation in local solar densities, bulge contribution, and extent of stars beyond the solar neighborhood. \texttt{m12f} has a fairly large tidal stream at 15-25 kpc that contributes to the enhancement at these distances, relative to the other two simulations (see Figure 3 of \citealt{Sanderson2017c}).}
\label{fig:disthist}
\end{figure*}

\subsection{Changing the synthetic surveys}
\label{subsec:reprocessing}

Users can alter the synthetic surveys as more knowledge about any of its components (the extinction model, the error model, or the selection function) is gained. The most complex task is to alter the extinction model, which implicitly requires reapplication of the other two components. In this section we outline how users can reprocess a synthetic survey under new assumptions.

Changing the extinction model will alter the apparent magnitudes and colors of the stars in the synthetic survey, requiring reapplication of the error model and selection function. To allow for changing the extinction model, we also provide the intermediate quantities $N_{H}^{\rm eff}$, $E(B-V)$, and $A_0$ used to compute the extincted magnitudes of each synthetic star, as well as their intrinsic magnitudes \super{G}{int}, \supersub{G}{Bp}{int}, \supersub{G}{Rp}{int} and the associated colors $\supersub{G}{Bp}{int}-\supersub{G}{Rp}{int}$, $\supersub{G}{Bp}{int}-\super{G}{int}$, and $\super{G}{int}-\supersub{G}{Rp}{int}$. Applying a new extinction model to the survey thus comprises the following four steps:
\begin{enumerate}
\item Use the new extinction model to recompute $E(B-V)$ and $A_0$ from $N_{H}^{\rm eff}$. Refer to \citet{2018arXiv180409378G} to convert these to extinctions $A_G$, $A_{G_{BP}}$, and $A_{G_{RP}}$ in the Gaia passbands.
\item Recompute new extincted apparent magnitudes from the intrinsic magnitudes provided by applying the new extinctions in each passband, for example:
\begin{equation}
\super{G}{ext} = \super{G}{int} + A_G
\end{equation}
If desired, also update the reddened colors.
\item Apply an error model (either the one in \S \ref{subsec:errormodel} or whatever is desired) to recompute the new observational uncertainties \textit{on all observed quantities}, since all depend on apparent magnitude. Re-sample from the unconvolved observed quantities using the new uncertainties for at least the apparent $G$ magnitudes, and for any other error-convolved quantities desired.
\item Apply a selection function (either the one in \S \ref{subsec:selection} or whatever is desired) to the \textit{error-convolved} apparent $G$ magnitudes to determine which stars now fall in the reprocessed synthetic survey.
\end{enumerate}

The ability to vary the extinction model \textit{post facto} will be limited by the fact that only stars whose extincted magnitudes \emph{in our dust model} are bright enough for \Gaia are included in the survey. Thus, extinction models that predict significantly {\em less} extinction than our adopted default will require re-running on the \textit{un-extincted} mock catalogs, rather than on the synthetic surveys. Since the un-extincted mocks contain 2--4 times as many stars as the extincted synthetic surveys, and we anticipate that most users will want to increase the extinction from our baseline and therefore will not need them, we have not made the un-extincted mocks available for public download but are happy to provide them on request.

Changing the error model or selection function requires only a subset of the steps enumerated here. If varying only the error model is desired, start from step 3 in the above list. To change the selection function, simply follow step 4.

\section{Results}
\label{sec:results}

To arrive at the nine synthetic \Gaia surveys in this data release, we generated a total of 125 trillion synthetic stars in the mock catalogs, of which around 43 trillion made it into the synthetic surveys after applying the extinction model and magnitude selection. Table \ref{tbl:files} summarizes the data release, which was split up into ten LSR-centric distance bins per survey for portability.
Because we ignore the effect of crowding and use a conservatively low value for the extinction, and because our simulated galaxies range in total stellar mass up to twice that of the MW, the synthetic surveys all have a larger total number of stars than the real \Gaia DR2.
% * <arwetzel@gmail.com> 2018-06-25T05:40:32.300Z:
% 
% these numbers should be billion, not trillion, right?
% 
% ^.

Figure~\ref{fig:disthist} shows a cumulative histogram of the stars in each synthetic survey as a function of their true LSR-centric distance, to illustrate how both the synthetic and real \Gaia surveys reach far beyond the local solar neighborhood. The wide variation in the contents of the synthetic surveys, especially in the outer reaches, underlines the importance of the fully cosmological context of our simulations. In this section we give an overview of the contents of the synthetic surveys.

\begin{figure*}
\begin{center}
\includegraphics[angle=90,height=0.9\textheight]{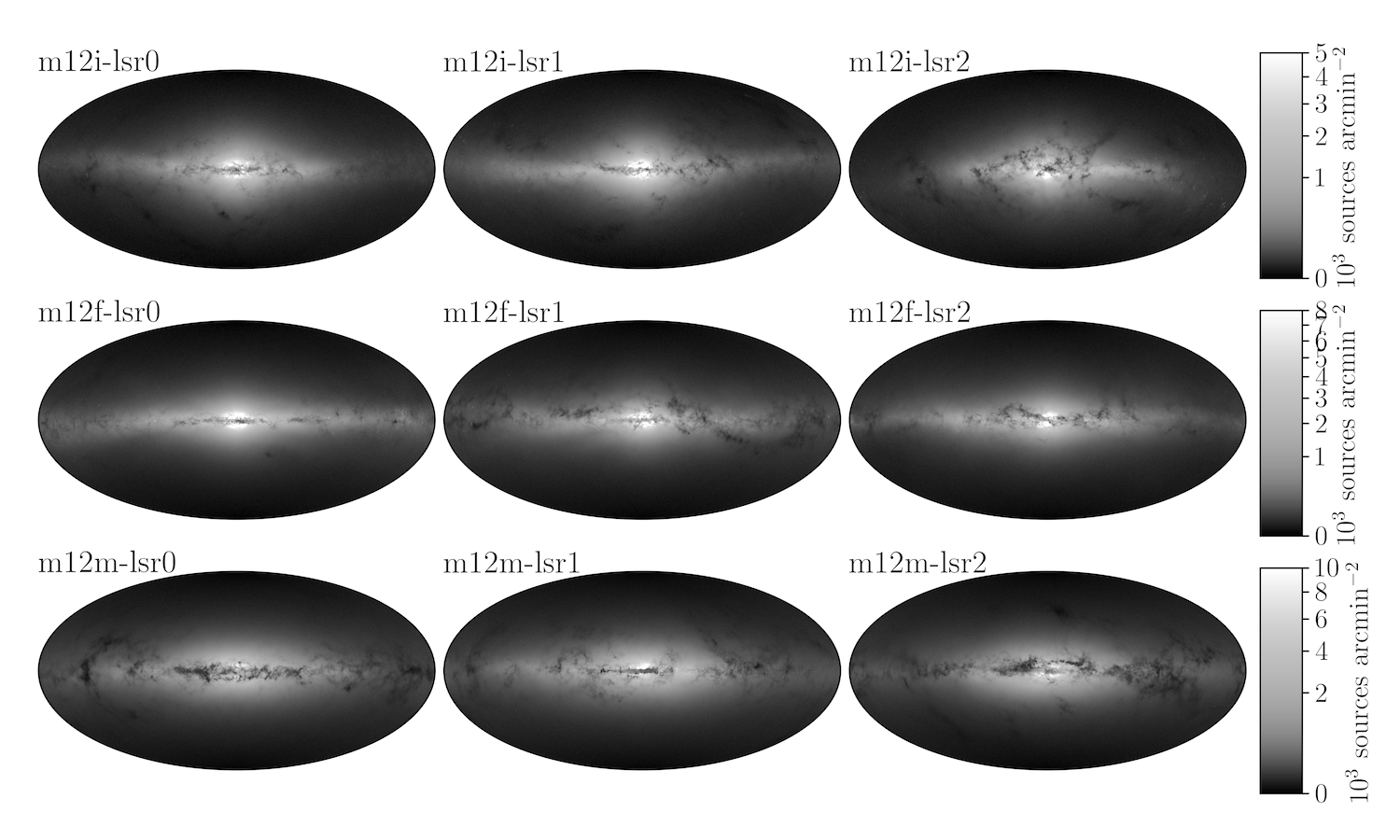}
\end{center}
\caption{All-sky star count maps of synthetic \textit{Gaia}-like surveys (compare to Figure~3 of \citealt{DR2}) display a wide variety of features, even for different viewpoints in the same galaxy. The labeling refers to the simulation and the solar viewpoint chosen; see Tables~\ref{tbl:solar-positions} and \ref{tbl:files}.
}
\label{fig:allsky}
\end{figure*}

\subsection{Multidimensional views of simulated galaxies}

One can compare our synthetic surveys directly with \Gaia DR2. Detailed comparisons of specific properties of the simulated versus MW populations are beyond the scope of this work; we discuss some aspects of a few such studies in progress in \S\ref{sec:future} but defer further comparison to future work. Instead we present a few views, generated from one of our synthetic surveys, that visually can be compared with some of the key first results from \Gaia DR2.

As for the real data, we generate the star-count maps shown in Figure~\ref{fig:allsky} (comparable to Figure~3 of \citealt{DR2}). The extinction distribution, which is prominent in our simulated maps as well as in the real \Gaia data, varies substantially even for different viewpoints within the same simulation. Some (like \texttt{m12f-lsr0} and \texttt{m12m-lsr1}) show a thin plane of dense extinction like the real MW, while in others (like \texttt{m12i-lsr2} and \texttt{m12f-lsr1}) the line-of-sight extinction has little or no identifiable thin structure and extends far out of the disk plane. When interpreting these views, it is important to keep in mind our assumption of a constant dust-to-metals ratio in computing the extinction (\S\ref{subsec:extinction}), and the generally higher gas masses of the simulated galaxies relative to the MW (Table~\ref{tbl:galaxies}).

Another notable difference between the simulated surveys and the MW is the absence of the Magellanic Clouds; none of the simulated galaxies have companions as large and close by. The galactic disk displays warps or truncations near anticenter in some cases (\texttt{m12i-lsr1}, \texttt{m12f-lsr0}), providing some interesting test cases for those interested in searching for such features in the MW. The bulge is fairly prominent in many cases behind the extinction, which serves as an important reminder that our synthetic surveys do not attempt to model crowding.

\begin{figure*}
\begin{center}
\includegraphics[height=0.3\textheight]{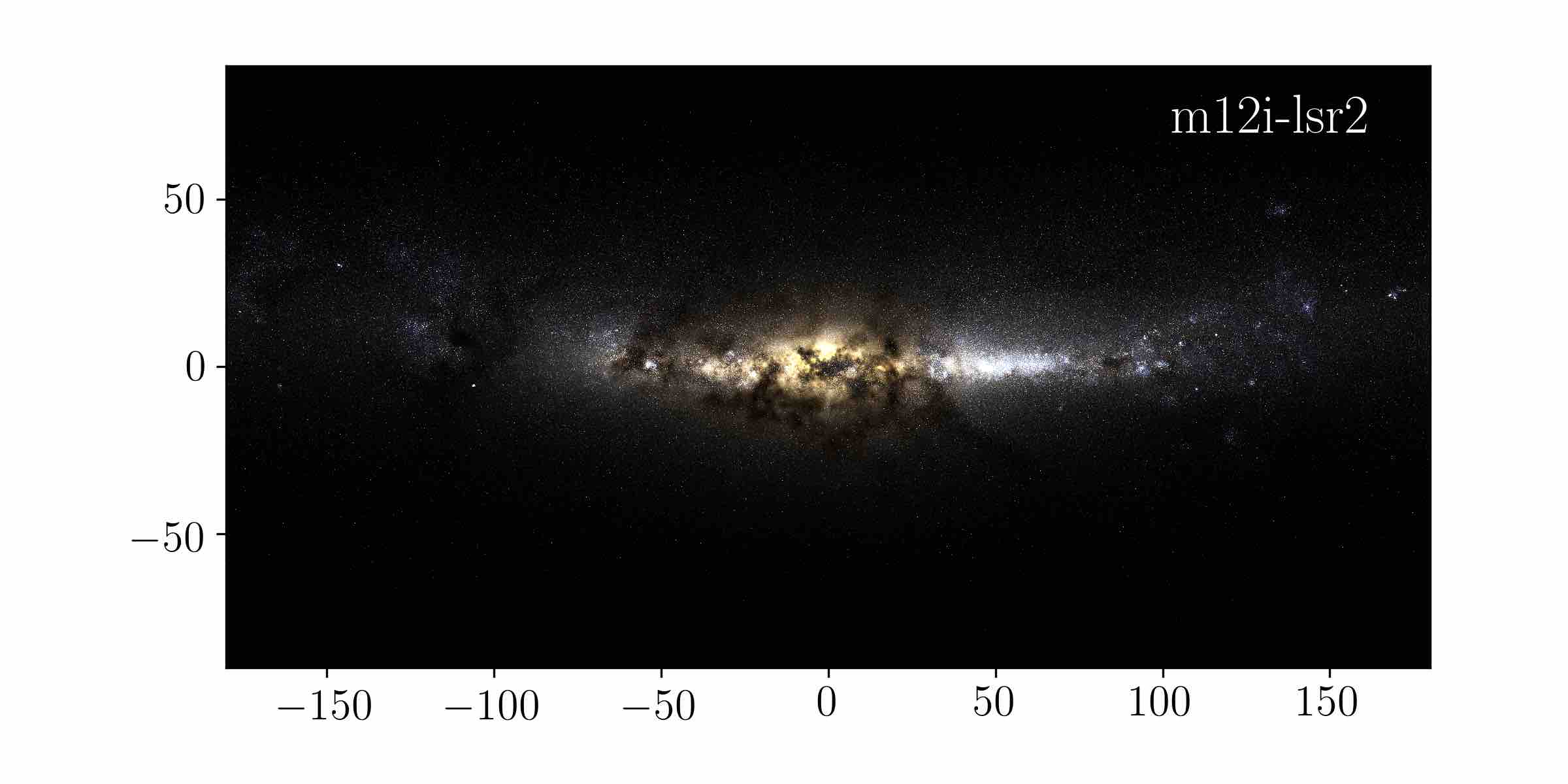}\\
\includegraphics[height=0.3\textheight]{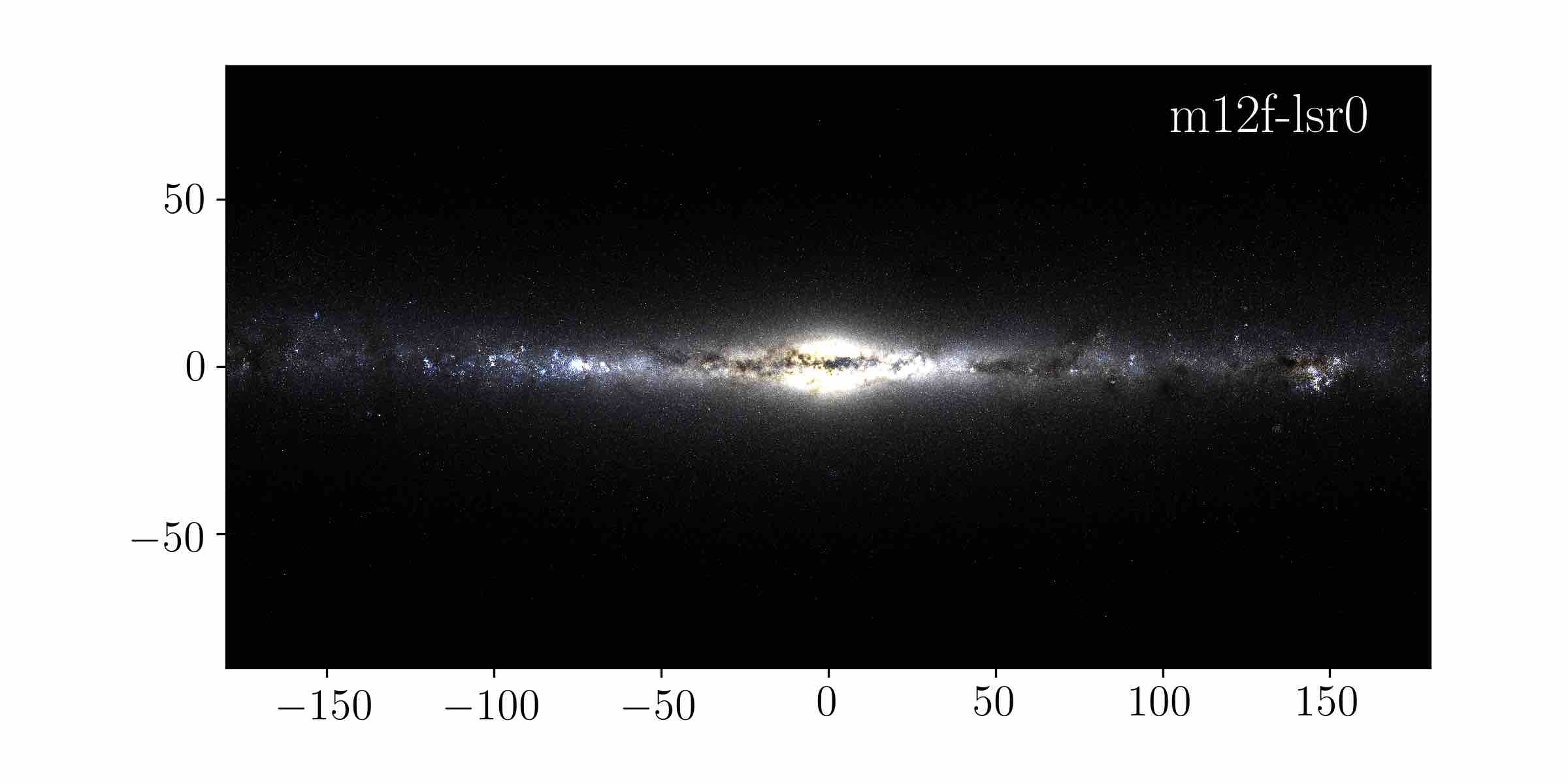}\\
\includegraphics[height=0.3\textheight]{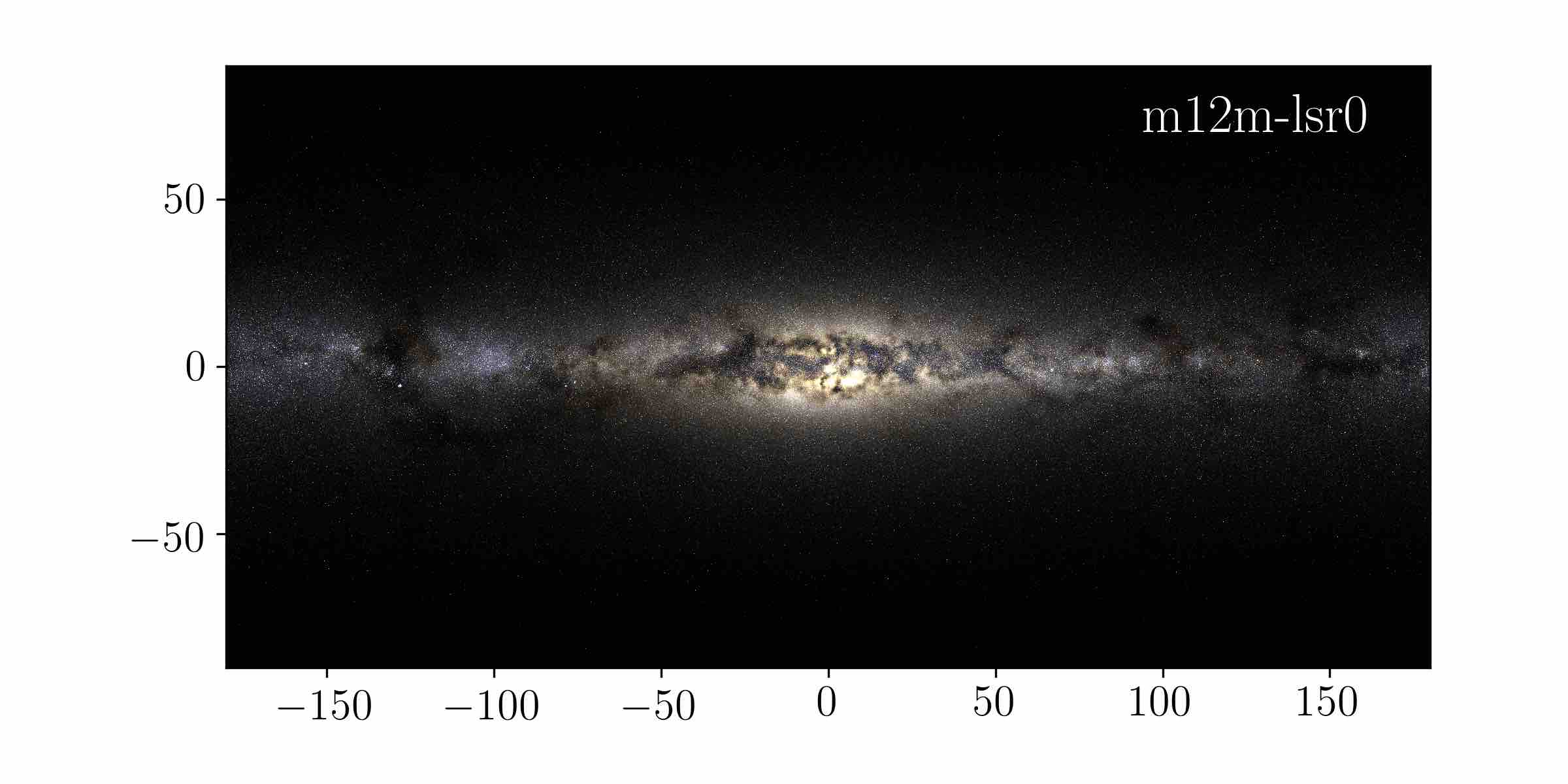}
\end{center}
\caption{RGB flux maps of three of the synthetic surveys, \texttt{m12i-lsr2} (top), \texttt{m12f-lsr0} (middle), and \texttt{m12m-lsr0} (bottom), highlight the differences in apparent color of different simulated galaxies. Total flux in the $G_{Rp}$ band was assigned to the $R$ channel, in the $G$ band to the $G$ channel, and in the $G_{Bp}$ band to the $B$ channel. The images were median-filtered using the 6 neighboring pixels to diminish the effect of ``hot'' pixels caused by individual nearby stars. Axes show galactic coordinates in degrees.
}
\label{fig:threecolor}
\end{figure*}

The variation in the synthetic surveys is also on display in Figure~\ref{fig:threecolor}, which shows RGB flux maps for one viewpoint from each of the three simulated galaxies (comparable to Figure~4 of \citealt{DR2}). Even the dominant colors of the different simulated galaxies vary substantially. Clusters of bright blue young stars are apparent in this view, highlighting the ability of our simulation code to resolve individual regions of star formation as well as the importance of a self-consistent extinction calculation.

\begin{figure*}
\includegraphics[width=\textwidth]{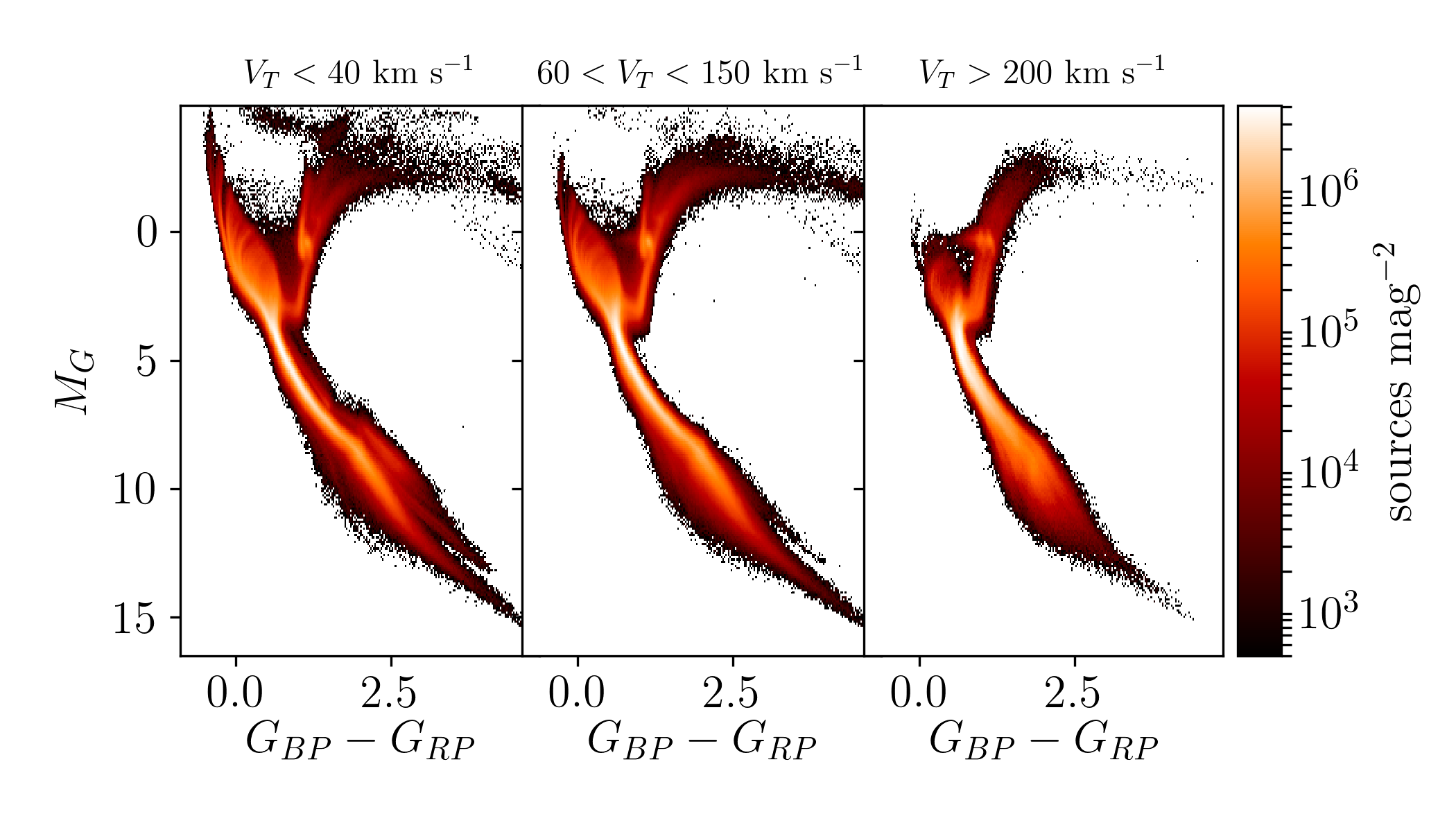}
\caption{The Hertzsprung-€"Russell diagram as a function of tangential velocity for stars in the synthetic survey \texttt{m12f-lsr0} shown in Figure~\ref{fig:allsky}, with quality selections as discussed in the text (compare Figure~21 of \citealt{2018arXiv180409378G}), showcases the variety in stellar populations of the ``solar neighborhood'' in the synthetic surveys. For this subsample of synthetic stars with negligible extinction and high-quality proper motions and distances, some echoes of the underlying grid of isochrones are still visible at the brightest magnitudes, where the model grid is sparsest. In less pristine subsamples these artifacts will not be apparent.}
\label{fig:hrd-vtan}
\end{figure*}

To illustrate the multi-dimensionality of the synthetic surveys, we show a set of three Hertzsprung-Russell diagrams (HRDs) of \texttt{m12m-lsr0} for different ranges of the transverse velocity $V_T$, expressed in terms of the proper motions $\mu_{\alpha*}$ and $\mu_{\delta}$ and the parallax $\varpi$ as
\begin{equation}
V_T \equiv \frac{1}{\varpi}\sqrt{\mu_{\alpha*} + \mu_{\delta}}.
\end{equation}
This strategy was adopted to create Figure~21 of \citet{2018arXiv180409378G} from the \Gaia DR2 data. As in that paper, we select for this plot only stars with estimated parallax error better than 10 percent, estimated $G$ magnitude error less than 0.22, estimated $G_{BP}$ and $G_{RP}$ errors less than 0.054, and extinction $A_{G} < 0.015$.  Selecting only stars with these accurate measurements means that some traces of the isochrone grid are still visible in the bright end of the HRDs where the grid is sparsest, but these blur out quickly if the data quality selections are loosened. Figure~\ref{fig:hrd-vtan} is qualitatively similar to what is seen in DR2 with the exception of the white dwarf sequence, which we do not simulate. Several separate main sequences can be identified in the stars moving with the LSR (left-hand panel), including one sequence that dominates for stars with larger $V_T$ (center panel). At highest $V_T$ (right-hand panel) the main sequence, the turnoff, and the red clump are significantly broader compared to the other panels, reflecting the heterogeneous mixture of stars at these velocities.

\begin{figure*}
\includegraphics[width=\textwidth]{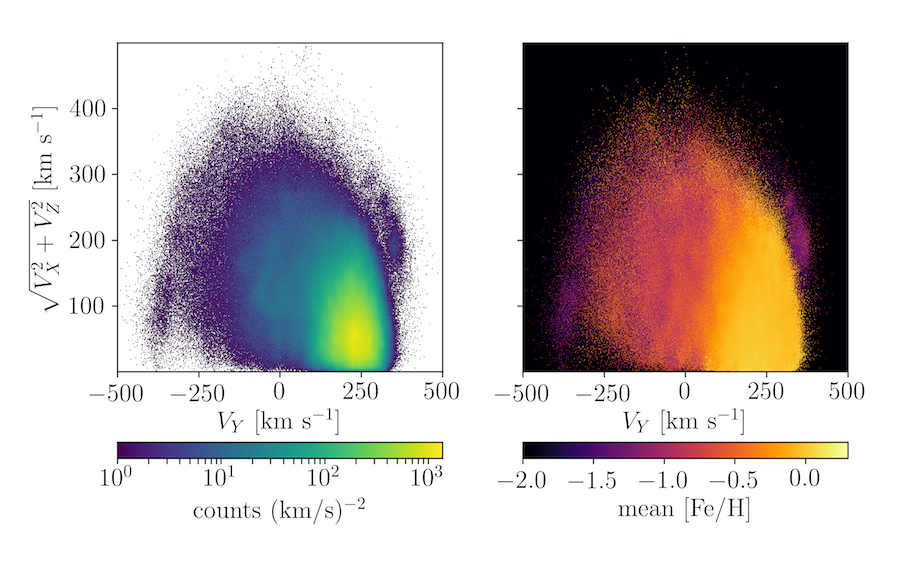}
\caption{Toomre diagram of \texttt{m12i-lsr0} in LSR coordinates (see \S\ref{subsec:heliocentric}) show the ability of the synthetic surveys to capture fine structure in a wide range of conditions. {\bf Left}: density of stars within 3 kpc of the LSR position that satisfy the quality cuts used in Figure~\ref{fig:hrd-vtan} and have measured radial velocities. \textbf{Right}: the same stars colored by mean [Fe/H] per pixel. Our high-fidelity sampling strategy has reproduced the small-scale clumpiness at high density near the LSR around (235,0) km \unit{s}{-1} and also captures stream-like structures at lower densities passing through the local ``solar volume'' near (-350,100) and (+350, 200) km \unit{s}{-1}. Viewed in abundance space (right), [Fe/H] varies with kinematic position as expected: stars moving with the LSR have solar-like [Fe/H] while the stream-like structures have lower [Fe/H] than even the smooth hot component.}
\label{fig:toomre}
\end{figure*}

As a second illustration of the power of synthetic phase-space surveys, we also present two views of the Toomre diagram of the LSR volume (stars within 3 kpc) for \texttt{m12i-lsr0} (Figure~\ref{fig:toomre}). In the right-hand panel is the standard view of the density distribution in the plane $(V_{\phi}, \sqrt{V_R^2 + V_z^2})$, where clustering and structure are apparent in both the stars moving with the LSR and in the diffuse, kinematically hot component. This illustrates the ability of our density sampling strategy to reproduce structures in velocity space at different scales and locations. We chose to show this example particularly because it includes examples of streams passing through the local volume on both highly retrograde and highly prograde orbits. The left-hand panel illustrates how adding abundance information gives clues to the identities of some of these features: the clusters near $V_{\phi}\sim 0$ belong to the local high-[Fe/H] disk, while the clumps near $(-350,\ 100)$ and $(+350,200)$ km \unit{s}{-1} have a lower [Fe/H] than even the rest of the hot component centered on the galactic center, supporting the idea that they are tidal streams intersecting the local volume.

\subsection{The effect of extinction on survey membership}

\begin{figure}
\begin{center}
\includegraphics[width=0.5 \textwidth]{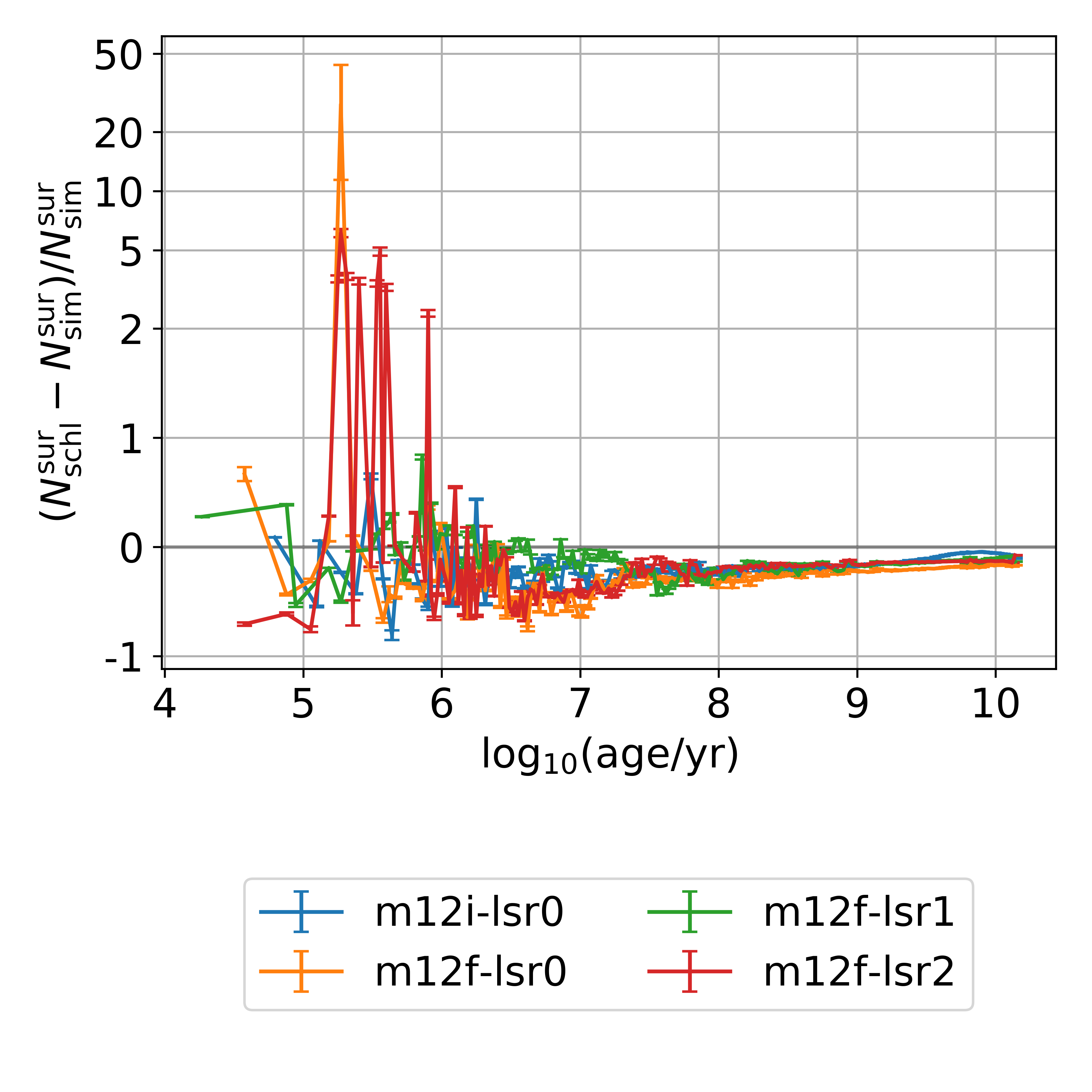}
\end{center}
\vspace{-1cm}
\caption{Fractional difference in numbers of stars in the final synthetic survey as a function of age, for different simulated galaxies and solar viewpoints, if the extinction is calculated based on the MW dust map (\supersub{N}{sur}{schl}) or on the simulated distribution of metal-rich gas ($\supersub{N}{sur}{sim}$). Error bars are propagated from Poisson uncertainties on the star counts in each age bin. Stars to the left of the vertical gray line are included in the contour plots in Figure \ref{fig:extinction_map}. For old stars, the number of stars entering the final survey is consistent across all viewpoints; surveys using the simulation to determine extinction include slightly more stars by design as discussed in \S \ref{subsec:extinction}. For the youngest (age $\lesssim1$ Myr) stars, the difference varies substantially from survey to survey (see discussion in text and Figure \ref{fig:extinction_map}), but applying the MW dust map generally results in overrepresentation of the youngest stars in the final synthetic survey.}

 %\textbf{Top}: density (in square-root scaling) of synthetic stars with ages younger than 100 Myr. \textbf{Middle:} $E(B-V)$ reddening (linear scale from 0 to 2), computed from the simulated gas metallicity distribution as described in \S\ref{subsec:extinction}, for a subset of the stars depicted in the top panel. \textbf{Bottom:} $E(B-V)$ reddening (linear scale from 0 to 2) computed from the \citet{drimmel} three-dimensional extinction map of the Milky Way. In the bottom two panels, stars at \emph{larger} distances, with higher reddening, are plotted on \emph{top} to emphasize the regions with the most dust. It is apparent from comparing the three panels that both proper correlations between young star clusters and regions of high extinction, and a correct global structure in the disk plane, require a self-consistent calculation of the extinction.
 
\label{fig:extinction_age}
\end{figure}

\begin{figure*}
\begin{center}
\includegraphics[width=\textwidth]{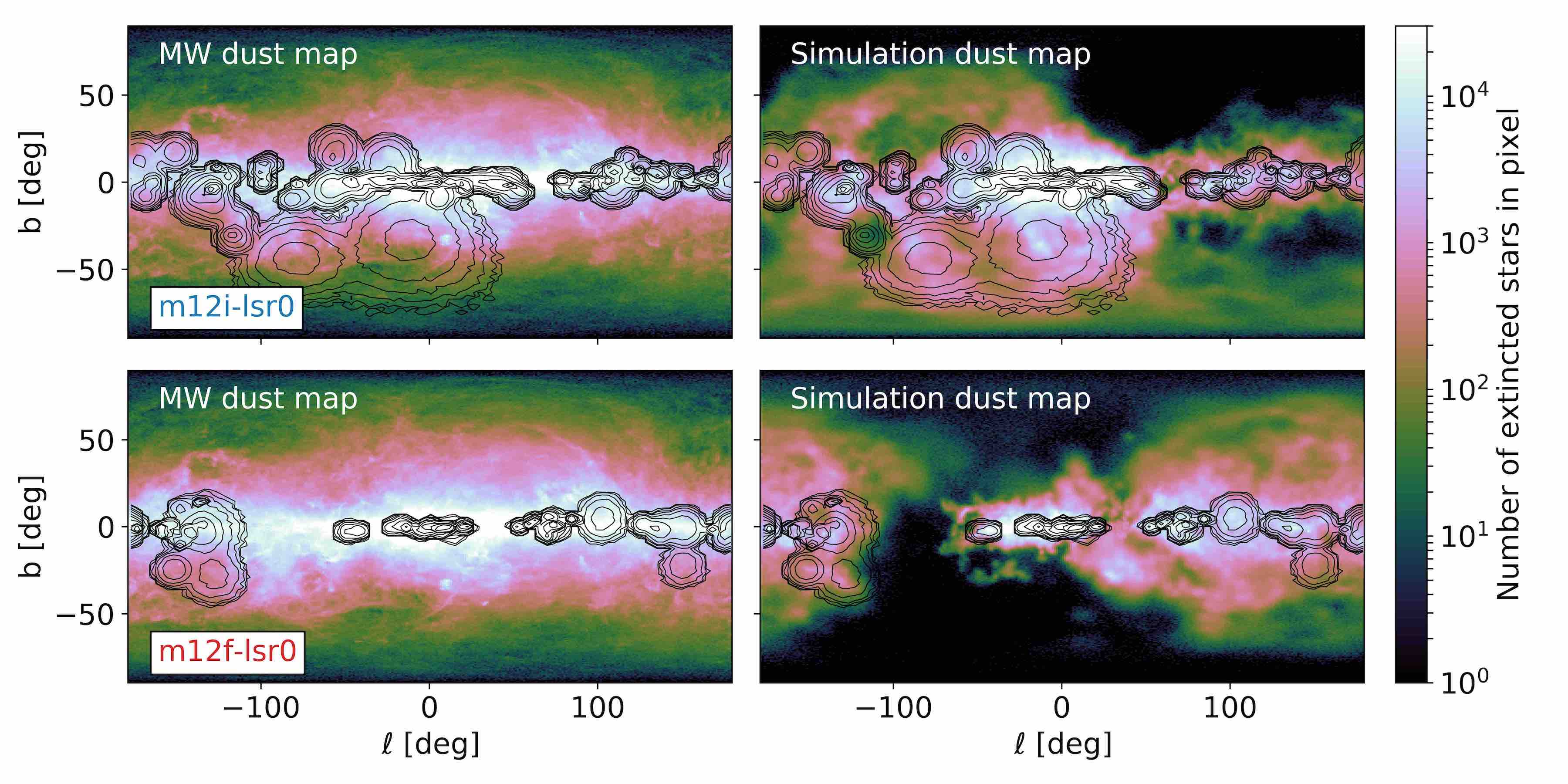}
\end{center}

\caption{Using the simulated gas density to estimate the dust extinction preserves the correlations between extinction and young (age $\lesssim 10$ Myr) stellar populations, which can differ significantly in their distribution from the MW. These examples show stars within 4.5 kpc of the LSR (``solar'') position. In each row, the panels show the number density of stars that do \textbf{not} end up in the final synthetic survey in color, for the MW dust map (left) and the map calculated from the simulation (right), overlaid with contours showing the projected density of young stars. The top row shows an example where the young stars are distributed more or less similarly to the regions of high extinction in the MW dust map (\texttt{m12i-0}; see Table \ref{tbl:solar-positions} and middle panel of Figure \ref{fig:threecolor}), while the bottom row shows an example where the distribution differs significantly from the MW (\texttt{m12f-0}).}

%\caption{Impact of different extinction models on quasi-local stellar populations ($\sub{d}{LSR}<4$ kpc) included in the synthetic survey \texttt{m12f-lsr0}. Left: HRD of all stars within the volume that could be included in the catalog (unextincted apparent $G$ magnitude $< 21$). Right: difference in the number of extincted stars (extincted apparent $G$ magnitude $> 21$) between the MW dust map and the extinction computed from the simulation. Younger stars are delineated by the two solar-metallicity isochrones with ages 10 Myr (black) and 100 Myr (magenta) in both panels. By design, the simulated extinction generally removes fewer stars from the survey than the MW map; however, the opposite is true for very young stars (above and to the right of the black isochrone) of which far fewer are removed by the MW map, in which the locations of dust clouds are uncorrelated with the starforming regions of the simulated galaxy.}

\label{fig:extinction_map}
\end{figure*}

Figures~\ref{fig:extinction_age} and \ref{fig:extinction_map} motivate our choice to calculate the extinction model for our synthetic surveys from the metal-enriched gas in the simulations rather than applying the observed MW dust map. Figure \ref{fig:extinction_age} shows the fractional difference in the number of stars ending up in the final survey as a function of age if the extinction is calculated based on the MW dust map (\supersub{N}{sur}{schl}) rather than the simulated distribution of metal-rich gas ($\supersub{N}{sur}{sim}$). For older stars (everything with age $\gtrsim 10$ Myr) the results are consistent across the different simulations and viewpoints, and reflect our deliberate choice to use a relatively low efficiency for our simulation-based dust maps, and therefore admit more stars into the final survey relative to the standard \citet{drimmel} MW dust map. For the youngest stars (age $\lesssim1$ Myr), however, the results vary quite significantly from simulation to simulation, and from viewpoint to viewpoint within the same simulated galaxy. In some cases (\texttt{m12f-lsr0}, \texttt{m12f-lsr2}) using the MW map results in a huge increase in the number of the youngest stars admitted to the survey, while in others (\texttt{m12i-lsr0}, \texttt{m12f-lsr1}) the relative fraction of the youngest stars is somewhat comparable to the relative fraction of older stars admitted.

Figure \ref{fig:extinction_map} illustrates the reason for these variations. The underlying maps show the number of \emph{extincted} stars as a function of sky position in Galactic coordinates in the same 4.5-kpc volume for the two different dust maps, with contours of the density of young (age $<10$ Myr) stars overlaid in black. In surveys that show a smaller difference in the number of young stars admitted between the two dust map choices, such as \texttt{m12i-lsr0} in the top row of the figure, the young stars tend to be distributed in a way that mirrors the regions of highest extinction in the MW dust map, which also more closely resembles the inferred extinction from the simulation itself. On the other hand, in surveys that show a large difference, such as \texttt{m12f-lsr0} in the bottom row of Figure \ref{fig:extinction_map}, the young stars are distributed quite differently from the high-extinction regions in the MW, yet trace closely the extinction inferred from the simulation. We can conclude from this that the extinction inferred from the simulation is better correlated with regions of ongoing star formation, and thus that too many young stars are being included in the synthetic surveys constructed using the MW dust map, rather than being extincted by the surrounding dust of their starforming regions. 

Some of the scatter in the youngest bins in Figure \ref{fig:extinction_age} could be ascribed to the fact that while there are many synthetic stars in each bin (at least 1000 in almost every bin), they are in some cases spawned from relatively few individual simulation star particles in those age bins (all stars spawned from the same particle have the same age) and mapped to the isochrone grid which also has finite age spacing. So some of the fluctuations in the number of extincted stars may have to do with these steps in the creation of the synthetic survey, which are unavoidably discretized at the resolution level of the simulation and the isochrone grid. This is the reason for choosing a relatively large volume over which to do the comparison, to ensure that the number of young star \emph{particles} is also large enough to get away from Poisson noise. The number of independent young star particles (ages younger than 1 Myr) used to spawn the young stars is approximately 80 for \texttt{m12f-lsr2} (which shows a large discrepancy between dust maps), $\sim40$ for \texttt{m12f-lsr0} (which also shows a high discrepancy), $\sim160$ for \texttt{m12f-lsr1} (less discrepancy), and $\sim65$ for \texttt{m12i-lsr0} (also less discrepancy). Given that especially \texttt{m12i-lsr0} has a comparable number of independent star particles to \texttt{m12f-lsr2} but shows a lower level of fluctuation between bins, we maintain that it is the consistency of the dust maps (as shown in Figure \ref{fig:extinction_map}) that is primarily determining the discrepancies we see in Figure \ref{fig:extinction_age}. The fluctuations in Figure \ref{fig:extinction_age} can thus be attributed to the fact that star formation is patchy in the simulations just as in real life, and so stars spawned from particles with different formation times are located in different places more or less masked by extinction in the final catalog.

Figure \ref{fig:extinction_map} also illustrates how the extinction calculated from the simulation fairly represents the sizes of resolved structures in the simulation. The observed extinction map of the MW itself can resolve smaller angular features in the dust distribution than the simulation is capable of representing at its current particle resolution. On the other hand, the resolution in line-of-sight distance of the MW map is significantly coarser than the simulation outside the immediate local volume. Applying the MW dust map properly to create a synthetic survey would therefore involve downgrading the angular resolution of the map everywhere to the local simulation resolution, to avoid introducing spurious small-scale features on top of the stellar distribution as realized by the simulation, while at the same time oversmoothing relative to the simulation resolution along the line of sight. Furthermore, we expect Gaia to permit construction of a far better 3D dust map of the Galaxy than is currently available. Rather than entangling all these current observational and numerical length scales, we simply chose to use the resolution-matched extinction inferred directly from the simulation where the local scale is well defined.

%Comparing the density distribution for young stars (black contours) to the extinction computed either by applying the \citet{drimmel} MW dust map (left) or from the simulated gas distribution (right), for a somewhat local volume ($\sub{d}{LSR}<4$ kpc) around the Sun, it is clear that a self-consistent calculation of the extinction---that is, one estimated from the simulated gas distribution---is required to properly represent the correlation between star-forming regions, young star clusters, and extinction in the simulated galaxy. When the total survey is depicted (as in Figure \ref{fig:threecolor}, middle panel) the disk extinction calculated this way appears relatively thin and well organized, similar to the MW's, but it is apparent that the distribution in the near field is quite different, further supporting the use of a self-consistent model. The difference between the details of the MW's dust distribution and the simulated galaxy's estimated one also supports our decision to estimate the extinction from the simulation rather than apply the MW dust map; as shown in Figure~\ref{fig:allsky}, some viewpoints have gas that extends (and extincts) far above and below the galactic plane or has a substantially different scale height than the MW (compare for example \texttt{m12i-lsr2} and \texttt{m12f-lsr0}, or even the different viewpoints of \texttt{m12m}; see the summary of scale height values in Table \ref{tbl:solar-circle}).

%Figure \ref{fig:extinction_map} shows another view of the impact of our choice of extinction model and dust efficiency, this time in its effect on the Herzsprung-Russell diagram (HRD). The entire HRD of all stars within 4 kpc that could be included in the catalog (unextincted apparent $G$ magnitude $< 21$) is shown in the left-hand panel, while the difference in the number of extincted stars (extincted apparent $G$ magnitude $> 21$) between the MW dust map and the extinction computed from the simulation is shown on the right. By design, our chosen model removes fewer stars than the MW map would have almost everywhere in the HRD. This implies that if a user wants to apply the MW dust map instead (despite the arguments above and in \S \ref{subsec:extinction}) they can do so using the data provided in the synthetic surveys (as described in \S \ref{subsec:reprocessing}) without missing stars. However, there are large swings in the difference between the two models especially for the youngest stars, to the right of the 10-Myr isochrone shown in black in both panels. This is another illustration of the importance of preserving correlations between the locations of young stars and regions of high extinction by using the gas distribution in the simulation to calculate the extinction.

\subsection{Data access}

As of this writing, the nine synthetic surveys described in this work are available through the data service \textsf{yt-hub}, at the website \url{https://ananke.hub.yt}. 
This service includes the ability to remotely analyze the synthetic surveys through a browser-based Jupyter notebook interface, eliminating the need to download the data to a local machine. In the near future they will also be available through the NASA/IPAC Infrared Science Archive, at \url{https://irsa.ipac.caltech.edu/Missions/ananke.html}, where they can be searched and explored alongside \Gaia DR2. We anticipate that this resource will expand to new simulations and new synthetic surveys in future and, given the size of the datasets, may move to a different dedicated site. An up-to-date list of data access points will be maintained at \url{http://fire.northwestern.edu/ananke}.

In addition to the nine synthetic surveys, we also provide alongside these the corresponding ``raw'' simulation snapshots of our three MW-mass systems at $z = 0$, including all dark matter, gas, and star particles.
As of this writing, we are releasing only each simulation's snapshot at $z = 0$, though in the future we plan to release higher-redshift snapshots as well.

\subsection{Contents and data model of the synthetic surveys}

In this section we describe in detail each of the fields in the synthetic surveys. We use the same names and units for quantities that are in the real DR2 catalog where this was suitable, and attempt to use a relatively consistent naming convention for other fields where possible.
As discussed in \S\ref{sec:synthetic-survey}, we report the intrinsic and error-convolved values for observables delivered by \Gaia: for a quantity Q, the field \texttt{Q} contains the error-convolved quantity, the field \texttt{Q\_true} contains the underlying intrinsic value and \texttt{Q\_error} contains the estimated error, such that \texttt{Q} is a single random sample from a Gaussian centered at \texttt{Q\_true} with standard deviation \texttt{Q\_error}.

The surveys each consist of a set of ten data files containing the synthetic stars for different ranges in distance. The distance ranges and the number of synthetic stars per file for each of the nine synthetic surveys are given in Table \ref{tbl:files}. The data model is summarized in Table~\ref{tbl:datamodel}, found at the end of this paper, in which quantities with names identical to \Gaia DR2 are listed first, followed by supplemental information.

\begin{deluxetable}{lll|rrr}
\tablecaption{Contents of the \textsf{ananke} synthetic surveys of the Latte MW-mass suite of FIRE simulations. \label{tbl:files}}

\tabletypesize{\scriptsize}
\tablecolumns{6}
\tablehead{
\\
\multicolumn{3}{c|}{{\footnotesize File information}} &
\multicolumn{3}{c}{{\footnotesize Number of stars per file}}
}
\startdata
 &  \sub{d}{min} & \sub{d}{max}  & \multicolumn{3}{c}{m12i}\\
 index & [kpc] & [kpc] & \multicolumn{1}{c}{lsr-0} & \multicolumn{1}{c}{lsr-1} & \multicolumn{1}{c}{lsr-2} \\
 \tableline
 0 & 0 & 3.0 & 317,251,458 & 358,924,610 & 392,096,021 \\
 1 & 3.0 & 4.25 & 281,083,240 & 290,094,002 & 289,405,931 \\
 2 & 4.25 & 5.5 & 377,105,602 & 397,004,385 & 345,162,879 \\
 3 & 5.5 & 6.5 & 363,758,333 & 413,609,097 & 323,184,686 \\
 4 & 6.5 & 7.25 & 319,295,239 & 387,632,445 & 273,828,420 \\
 5 & 7.25 & 8.0 & 350,782,201 & 450,147,130 & 295,219,748 \\
 6 & 8.0 & 9.0 & 412,716,462 & 529,250,734 & 341,842,225 \\
 7 & 9.0 & 10 & 262,217,693 & 320,513,185 & 212,660,571 \\
 8 & 10 & 15 & 418,391,956 & 502,481,997 & 361,540,407 \\
 9 & 15 & 300 & 112,963,541 & 104,844,392 & 97,221,224 \\
  \tableline
 \multicolumn{3}{r|}{Total} & 3,215,565,725 & 3,754,501,977 & 2,932,162,112 \\
   \tableline \tableline
 &  \sub{d}{min} & \sub{d}{max}  & \multicolumn{3}{c}{m12f} \\
  index & [kpc] & [kpc] & \multicolumn{1}{c}{lsr-0} & \multicolumn{1}{c}{lsr-1} & \multicolumn{1}{c}{lsr-2} \\
   \tableline
 0 & 0 & 3.0 & 430,683,038 & 605,353,233 & 550,603,568 \\
 1 & 3.0 & 4.25 & 397,435,584 & 405,405,662 & 422,441,994 \\
 2 & 4.25 & 5.5 & 564,297,458 & 493,908,824 & 480,957,890 \\
 3 & 5.5 & 6.5 & 575,873,595 & 470,547,860 & 444,786,876 \\
 4 & 6.5 & 7.25 & 543,929,978 & 414,946,045 & 387,745,610 \\
 5 & 7.25 & 8.0 & 644,919,694 & 493,714,456 & 451,382,016 \\
 6 & 8.0 & 9.0 & 777,622,811 & 605,746,675 & 551,845,344 \\
 7 & 9.0 & 10 & 491,535,388 & 338,956,912 & 332,524,725 \\
 8 & 10 & 15 & 975,639,189 & 586,512,307 & 654,192,010 \\
 9 & 15 & 300 & 449,470,541 & 291,448,782 & 402,362,139 \\
  \tableline
 \multicolumn{3}{r|}{Total} & 5,851,407,276 & 4,706,540,756 & 4,678,842,172 \\
   \tableline \tableline
 &  \sub{d}{min} & \sub{d}{max}  & \multicolumn{3}{c}{m12m} \\
 index & [kpc] & [kpc]  & \multicolumn{1}{c}{lsr-0} & \multicolumn{1}{c}{lsr-1} & \multicolumn{1}{c}{lsr-2} \\
 \tableline
0 & 0 & 3.0 & 985,616,757 & 1,076,263,904 & 902,177,868 \\
 1 & 3.0 & 4.25 & 698,130,518 & 765,224,083 & 653,165,049 \\
 2 & 4.25 & 5.5 & 765,352,276 & 809,366,609 & 741,318,984 \\
 3 & 5.5 & 6.5 & 628,267,662 & 659,742,282 & 622,777,751 \\
 4 & 6.5 & 7.25 & 470,337,765 & 495,954,561 & 465,535,396 \\
 5 & 7.25 & 8.0 & 450,067,060 & 474,301,928 & 450,597,589 \\
 6 & 8.0 & 9.0 & 500,687,793 & 1,017,982,364 & 506,903,256\\
 7 & 9.0 & 10 & 350,820,320 & 379,507,637 & 351,237,158 \\
 8 & 10 & 15 & 673,146,968 & 737,331,255 & 662,909,637 \\
 9 & 15 & 300 & 179,332,262 & 180,165,772 & 160,212,422 \\
 \tableline
 \multicolumn{3}{r|}{Total} & 5,701,759,381 & 6,595,840,395 & 5,516,835,110 \\
 \enddata
\tablecomments{``Index'' is the file number containing information for synthetic stars in the distance range \sub{d}{min}--\sub{d}{max}. $N_*$ is the number of synthetic stars in each distance range. Table~\ref{tbl:solar-positions} gives the solar viewpoint locations for the different surveys.}
\end{deluxetable}

\subsubsection{Indices}

We provide a number of different integers for identifying synthetic stars and associating them with their generating star particle in the simulation snapshots.
\begin{description}
\item[source\_id] is a unique integer for each synthetic star in a given synthetic survey (IDs may be reused in different surveys but do not indicate an association).
\item[random\_index] can be used to select a random subset of synthetic stars \emph{within the distance range of one data file}. To select a random subset of $N$ stars from one file, sort by \texttt{random\_index} and select the first $N$. To generate a random subset from an entire survey broken up over separate files, one should choose $n_i$ stars from each of the data files comprising the survey, where $\sum n_i = N$ and the individual $n_i$ are chosen by Poisson distribution based on the number of stars per file (given in Table \ref{tbl:files}).
\item[parentid] is the array index of the star particle from which a given synthetic star was spawned. Synthetic stars with the same \texttt{parentid} came from the same star particle, and the index can be used to locate the full properties of the star particle in the $z = 0$ snapshot. It is also useful to get a sense of the local scale of the smoothed kernel used to distribute synthetic stars in phase space, by plotting all synthetic stars with a common \texttt{parentid} in a region of interest.
Note that star particles in the FIRE simulations have an ID associated with them; however, our pipeline never uses this ID (because it is not always unique across star particles). Thus, we emphasize that \texttt{parentid} refers to array \textit{index} of the star particle within the simulation snapshot file(s).
\item[partid] indicates whether a synthetic star is located at the exact phase-space coordinates of its generating star particle. The first synthetic star to be spawned (\texttt{partid}=0) is always assigned the coordinates of the generating particle, but this star may or may not fall in the survey volume, even if others drawn from the local phase-space kernel around the same generating particle (\texttt{partid}=1) do. This often provides a quick way to access some information about the generating star particle without loading the complete snapshot.
\end{description}

\subsubsection{Phase space}

The heart of the real and mock \Gaia surveys is the position and velocity information for each star in the survey. We start with Cartesian positions and velocities in the LSR frame, and use the transformations provided by the \textsf{astropy} package \citep{astropy} to calculate positions and proper motions on the celestial sphere (right ascension and declination) and in Galactic latitude and longitude. In order to transform to Galactic coordinates, we rotate each system by the angle
\begin{equation}
\phi = \pi + \arctan\left(\frac{y_{\mathrm{LSR}}}{x_{\mathrm{LSR}}}\right)
\end{equation}
about the $Z$ axis, to follow the standard convention that the line from the LSR position to the Galactic center points in the positive $\hat{x}$ direction (the ``sun'' is on the $-X$ axis), and that the solar rotation is in the $+\hat{y}$ direction. We use the rotated stellar positions to compute $\ell$, $b$, and their associated proper motions. From these we then make the transformation to celestial-sphere coordinates. Given these assumptions, Table~\ref{tbl:datamodel}  provides the types of phase-space coordinates listed below.
\begin{description}
\item[px\_true, py\_true, pz\_true] are the intrinsic or ``true'' Cartesian position of the synthetic star relative to the LSR, placing the Sun at $x = -8.2$ kpc. To get back to the galactocentric principal-axis frame described in \S\ref{sec:coords}, first rotate the positions and velocities in the $x$--$y$ plane by the angle $\phi$ based on the values in Table \ref{tbl:solar-positions}, then use Equations~\ref{eq:gc1}--\ref{eq:gc2} to translate back to the galactocentric frame.
\item[vx\_true, vy\_true, vz\_true] are the true Cartesian velocity of the synthetic star relative to the LSR, related to the galactocentric principal-axis frame in the same way as the positions.
\item[dhel\_true, dmod\_true] are the LSR-centric distance and distance modulus of the synthetic star, related by 
\begin{equation}
\mathrm{dmod}_{\mathrm{true}}=5\log_{10}(d_{\mathrm{hel},\mathrm{true}})-5.
\end{equation}
\item[parallax\_true, parallax] are the true and error-convolved parallaxes of the synthetic star, with the associated estimated error \texttt{parallax\_error} from the model described in \S\ref{sec:synthetic-survey} and Table~\ref{tbl:errormodel}. The parallax is related to the LSR-centric distance via
\begin{equation}
\frac{\mathrm{parallax}_{\mathrm{true}}}{\mathrm{mas}} = \frac{1.0 \mathrm{kpc}}{d_{\mathrm{hel},\mathrm{true}}}.
\end{equation}
\item[ra\_true, ra; dec\_true, dec] are the true and error-convolved positions of the synthetic star on the celestial sphere, with their associated estimated errors \texttt{ra\_error} and \texttt{dec\_error}. The true RA and dec are calculated from the true Galactic latitude and longitude (themselves a product of the LSR-centric Cartesian coordinates) before error convolution.
\item[pmra\_true, pmra; pmdec\_true, pmdec] are the true and error-convolved proper motions of the synthetic star in the RA and declination directions, with their associated estimated errors \texttt{pmra\_error} and \texttt{pmdec\_error}. The RA proper motion includes the standard factor $\cos(\texttt{dec})$.
\item[radial\_velocity\_true, radial\_velocity] are the true and error-convolved radial velocities (with estimated error \texttt{radial\_velocity\_error}). \Gaia measures radial velocities for stars down to a brighter limiting magnitude than the astrometric survey. We report \texttt{radial\_velocity\_true} for all stars, but provide \texttt{radial\_velocity} and \texttt{radial\_velocity\_error} only for stars satisfying the magnitude and temperature limits cited by \Gaia DR2, as discussed in \S \ref{sec:synthetic-survey}.
\item[l\_true, l; b\_true, b] are the Galactic longitude and latitude of the synthetic star. The ``true'' values are transformed from the true Cartesian positions while the error-convolved values are transformed from the error-convolved RA and dec.
\end{description}

\subsubsection{Photometry}

We provide three types of photometric quantities in the synthetic surveys: 
\begin{description}
\item[\textbf{\textrm{intrinsic}}] magnitudes and colors (suffix \texttt{\_int}), which do not include any extinction or reddening and are not error-convolved;
\item[\textbf{\textrm{true}}] magnitudes and colors (suffix \texttt{\_true}), which have an extinction model applied to each band (hence the magnitudes are extincted and colors are reddened) but have not been error-convolved;
\item[\textbf{\textrm{observed}}] magnitudes and colors (no suffix), which have been extincted and then error-convolved according to the error estimated from the \texttt{true} extincted magnitudes. The \texttt{\_error} used to generate each observed magnitude is also provided.
\end{description}
We also provide a series of other quantities related to the extinction calculation, which can be used to change the extinction model if desired. 
\begin{description}
\item[lognh] is the effective metal-weighted column density of hydrogen along the line of sight between the synthetic star and the solar position, as defined in \S \ref{subsec:extinction}.
\item[ebv] is the reddening given $N_{H}^{\rm eff}$ above, related by the dust efficiency coefficient as discussed in \S \ref{subsec:extinction}.
\item[A0] is the extinction at 550 nm, related to the reddening above via the standard dust model $A_0 = 3.1 E(B-V)$.
\item[a\_g\_val, a\_bp\_val, a\_rp\_val] are the line-of-sight extinctions in the \Gaia $G$, $G_{BP}$, and $G_{RP}$ bands, calculated from the extinction at 550 nm ($A_0$) using the polynomial models in \citet{}, as described in \S \ref{subsec:extinction}.
\item[e\_bp\_min\_rp\_val] is the reddening between the two \Gaia spectrophotometric bands, $E(G_{BP}-G_{RP})$, equivalent to \texttt{a\_bp\_val}$-$\texttt{a\_rp\_val}.
\end{description}

\subsubsection{Stellar parameters}

We provide unconvolved values of stellar parameters for the synthetic stars in our catalog, interpolated from the isochrone models described in \ref{sec:isochrones}. \Gaia DR2 reports values without 1D uncertainties for some of these quantities, because the error model is fairly complex; the derivation of these parameters is expected to improve substantially in subsequent data releases. We therefore do not attempt to simulate any observational errors on these values.
\begin{description}
\item[teff\_val] is the effective temperature as given by the isochrone model.
\item[lum\_val] is the stellar total luminosity.
\item[logg] is the log surface gravity in cm \unit{s}{-2}.
\item[mact] is the present-day stellar mass of the synthetic star, accounting for stellar evolution.
\item[mini] was the mass of the synthetic star on the zero-age main sequence.
\item[age] is the log of the stellar age in Gyr, passed directly from the generating star particle and used to select the stellar isochrone to represent the single stellar population. All stars spawned from the same generating particle have the same age.
\item[mtip] is the mass of a star with the same age and metallicity at the tip of the giant branch. Evolved stars can be simply selected using the criterion \texttt{mact}$>$\texttt{mtip}. All stars spawned from the same generating particle share this value.
\end{description}

\subsubsection{Abundances}

FIRE simulations track 11 elemental abundances (H, He, C, N, O, Ne, Mg, Si, S, Ca, Fe) through the stellar yield table network and subgrid turbulent metal diffusion models described in \S\ref{subsec:numerics}. We pass these elemental abundances directly to each synthetic star from its generating star particle, consistent with our assumption of single-aged, single-abundance stellar population. Thus all stars sharing the same generating particle will have the same abundances. All abundances are reported compared to hydrogen and relative to the solar abundance (for which we take the values reported in \citealt{asplund09}, consistent with the values assumed to map to the model isochrones). We also provide for convenience the quantity \texttt{alpha}, which is the ratio of magnesium to iron abundance relative to solar. The full list of abundances is provided in Table \ref{tbl:datamodel}.

\section{Using the synthetic surveys}
\label{sec:quickstart}

Below are a few points to keep in mind when starting out using the synthetic surveys.  
\begin{itemize}
\item Scripts that run on Gaia DR2 data should work without much tweaking on the synthetic surveys, because \textbf{fields common to the synthetic and real surveys have identical names}.
There are two main exceptions:
\begin{itemize}
\item In the synthetic surveys \textbf{\texttt{ra\_error} and \texttt{dec\_error} are in degrees} (consistent with the units of \texttt{ra} and \texttt{dec}), not milliarcsec (as in DR2).
\item Some data quality flags are not present in the mock data, because there was no way to model them; columns in both the mock and real data are listed first in 
 Table~\ref{tbl:datamodel}.
\end{itemize}
\item Depending on the science case, not all ten files associated with each synthetic survey may be needed, so keep the distance bins (Table~\ref{tbl:files}) in mind. For example, all stars with complete 6-D positions and good parallaxes (sufficient for using distance=1/parallax) are in slices 0 and 1 of each synthetic survey.
\item As discussed in \S \S \ref{subsec:extinction} and \ref{sec:results}, and illustrated in Figure \ref{fig:extinction_age}, \textbf{the extinction model for the synthetic surveys is deliberately conservative}, using a value for the dust efficiency that is on the low end of the range estimated for the MW. As a result the mocks all contain more stars than the real DR2 even when the local stellar density is comparable. If users want to assume a stronger extinction model, the parameter $Q_{\rm dust}$ in Equation~\ref{eq:reddening} can be increased; a prescription for how to reprocess the survey with a different $Q_{\rm dust}$ is given in \S \ref{subsec:reprocessing}. Likewise if an analysis assumes the MW extinction map to correct for the selection function when processing the mock data, this will yield incorrect results since the extinction in the synthetic surveys is computed from the distribution of metal-enriched gas in the simulation (and is therefore different in its spatial distribution from the MW). If a user desires to overlay the MW dust map rather than the internally consistent extinction, they can do so according to the directions in \S \ref{subsec:reprocessing}. 
\item If an analysis is particularly sensitive to the details of the DR2 selection function, a parallel calculation of---and correction for---the selection function should be applied to the mock data before using them, keeping in mind the differences in the extinction map between these synthetic surveys and the MW. There are several options available publicly depending on the science case.
\item There are two ways to access information about the star particle that generated a particular synthetic star in a survey:
\begin{itemize}
\item The column \texttt{parentid} contains the \textit{index} of the generating star particle; to access it in the corresponding simulation snapshot, read in the entire snapshot and then use this value to index into the arrays. For example, if the snapshot is read into the object \texttt{part}, then the star particle's position is \texttt{part}[\lq star\rq][\lq position\rq][\texttt{parentid}]. (Note that star particles in the simulation snapshot have an ID parameter, but we do not use this in creating the synthetic catalogs, so users should ignore it.)
\item The column \texttt{partid} has the value 0 for synthetic stars that have the exact position and velocity of their generating particle, so this information can be accessed without loading the simulation snapshot by selecting a star with the same \texttt{parentid} as the one of interest and \texttt{partid} equal to 0. Properties like the chemical abundances and ages of synthetic stars are carried over identically from the generating particle as well; these properties are noted in Table \ref{tbl:datamodel}.
\end{itemize}
\item Finally: remember that \textbf{our FIRE cosmological simulations are not the Milky Way}: if an analysis is tailored to the MW's structure too specifically, it is likely to fail on the synthetic surveys. Either use the snapshots and visualizations bundled with the synthetic surveys to select the ones that best match the assumptions of the analysis method, or relax the assumptions. Conversely, these synthetic surveys provide a framework for testing whether a given inference on the observed MW is robust to effects (and their uncertainties) such as detailed morphology, structure, and dynamical state.
\end{itemize}

\section{Future work} 
\label{sec:future}

We anticipate that we and other members of the community will make detailed comparisons between these simulations and \Gaia for many different science cases, including the evolutionary structure of the disk and the phase-space structure of accreted stars. We plan to explore further, for example, the reason for the differences in the age-velocity dispersion relation observed in Figure~\ref{fig:sigma}. We also plan to validate statistical methods for modeling the Galactic mass distribution (e.g., \citealt{2015ApJ...801...98S}) using these catalogs to realistically represent both the expected observational uncertainties and a fully cosmological galaxy.

Currently, the FIRE project has $\sim 15$ simulations of MW-mass galaxies at sufficient resolution that could be added to this initial database. As with the real \Gaia dataset, we anticipate that periodic future releases will incorporate synthetic surveys of these new simulations. Users should visit \url{http://fire.northwestern.edu/ananke} for an up-to-date list. Eventually, we plan to release a public version of \textsf{ananke} alongside a webtool for creating user-described synthetic surveys of simulation snapshots. We hope that the tools provided here and in these subsequent releases will prove a fitting counterpart to a new, data-rich era of Milky Way science.

\section*{Acknowledgments}

The authors thank Justin Howell and Vandana Desai of IPAC, Kacper Kowalik and Matt Turk of yt, and Mark Bartelt at Caltech for their crucial assistance with the public data releases. We thank Anthony Brown for discussions on the characteristics of Gaia DR2 and Julianne Dalcanton for advice on models of dust extinction.

This work grew out of two series of \Gaia preparatory meetings focused on data analysis challenges. First, the \Gaia Challenge Workshops (held 2011--2015), which were organized through the \Gaia Research for European Astronomy Training Initial Training Network programme supported by the European Commission through its FP7 Marie Curie programme under grant agreement 264895. Second, the \Gaia Sprints (held 2016--present). Code for this project was developed in part at the 2017 Heidelberg \Gaia Sprint, hosted by the Max-Planck-Institut f\"ur Astronomie, Heidelberg.

This work has made use of data from the European Space Agency (ESA) mission Gaia (\url{http://www.cosmos.esa.int/gaia}), processed by the Gaia Data Processing and Analysis Consortium (DPAC, \url{http://www.cosmos.esa.int/web/gaia/dpac/consortium}). Funding for the DPAC has been provided by national institutions, in particular the institutions participating in the Gaia Multilateral Agreement.

RES was supported by an NSF Astronomy \& Astrophysics Postdoctoral Fellowship under grant AST-1400989, and by NASA through grant JPL 1589742.
AW was supported by NASA through ATP grant 80NSSC18K1097 and grants HST-GO-14734 and HST-AR-15057 via STScI.
Support for SL was provided by NASA through Hubble Fellowship grant HST-JF2-51395.001-A awarded by the Space Telescope Science Institute, which is operated by the Association of Universities for Research in Astronomy, Inc., for NASA, under contract NAS5-26555.
Support for PFH was provided by an Alfred P. Sloan Research Fellowship, NSF Collaborative Research Grant \#1715847 and CAREER grant \#1455342, and NASA grants NNX15AT06G, and JPL 1589742.
Support for SGK was provided by NASA through Einstein Postdoctoral Fellowship grant number PF5-160136 awarded by the Chandra X-ray Center, which is operated by the Smithsonian Astrophysical Observatory for NASA under contract NAS8-03060.
CAFG was supported by NSF through grants AST-1412836, AST-1517491, AST-1715216, and CAREER award AST-1652522, by NASA through grant NNX15AB22G, and by a Cottrell Scholar Award from the Research Corporation for Science Advancement.
DK was supported by NSF grant AST-1715101 and the Cottrell Scholar Award from the Research Corporation for Science Advancement.
EQ was supported by a Simons Investigator Award from the Simons Foundation and by NSF grant AST-1715070.

Numerical calculations were run on the Caltech compute cluster ``Wheeler,'' allocations from XSEDE TG-AST130039 and PRAC NSF.1713353 supported by the NSF, NASA HEC SMD-16-7592, and the High Performance Computing at Los Alamos National Lab.


\software{matplotlib \citep{matplotlib}, scipy \citep{scipy}, astropy \citep{astropy}, qhull \citep{qhull},
galaxia \citep{sharma11}, enlink \citep{sharmaEnlink}.}



%\appendix

\begin{longrotatetable}
\begin{deluxetable*}{lp{4in}ll}
\tabletypesize{\footnotesize}
\tablecolumns{4}
\tablecaption{Data model for synthetic surveys\label{tbl:datamodel}}
\tablehead{
\colhead{Quantity} &
\colhead{Explanation} &
\colhead{Data type} &
\colhead{Unit} 
}
\startdata
\multicolumn4c{\textbf{Fields with names identical to those in DR2}} \\
\hline
\hline
\multicolumn4c{Indices} \\
\hline
\texttt{source\_id} & Unique source identifier (per mock catalog) & long & \nodata  \\
\texttt{random\_index} & Random index used to select subsets & long & \nodata \\
\hline
\multicolumn4c{Astrometry} \\
\hline
\texttt{ra} & Right ascension & double & Angle (deg) \\
\texttt{ra\_error} & Standard error of right ascension & double & Angle (deg) \\
\texttt{dec} & Declination & double & Angle (deg) \\
\texttt{dec\_error} & Standard error of declination & double & Angle (deg) \\
\texttt{parallax} & Parallax & double & Angle (mas)  \\
\texttt{parallax\_error} & Standard error of parallax & double & Angle (mas)  \\
\texttt{parallax\_over\_error} & Parallax divided by its error & float & \nodata \\
\texttt{pmra} & Proper motion in RA direction & double & Angular Velocity (mas/year) \\
\texttt{pmra\_error} & Standard error of proper motion in RA direction & double & Angular Velocity (mas/year)  \\
\texttt{pmdec} & Proper motion in declination direction & double & Angular Velocity (mas/year)  \\
\texttt{pmdec\_error} & Standard error of proper motion in declination direction & double & Angular Velocity (mas/year)  \\
\texttt{l} & Galactic longitude (converted from RA, dec) & double & Angle (deg) \\
\texttt{b} & Galactic latitude (converted from RA, dec) & double & Angle (deg) \\
\hline
\multicolumn{4}{c}{Photometry}\\
\hline
\texttt{phot\_g\_mean\_mag} & \textbf{Extincted} apparent $G$-band mean magnitude  & float & Magnitude (mag) \\
\texttt{phot\_g\_mean\_mag\_error} & Standard error of $G$-band mean magnitude  & float & Magnitude (mag) \\
\texttt{phot\_bp\_mean\_mag} & \textbf{Extincted} apparent $G_{Bp}$-band mean magnitude  & float & Magnitude (mag) \\
\texttt{phot\_bp\_mean\_mag\_error} & Standard error of $G_{Bp}$-band mean magnitude  & float & Magnitude (mag) \\
\texttt{phot\_rp\_mean\_mag} & \textbf{Extincted} apparent $G_Rp$ band mean magnitude  & float & Magnitude (mag) \\
\texttt{phot\_rp\_mean\_mag\_error} & Standard error of $G_{Rp}$-band mean magnitude  & float & Magnitude (mag) \\
\texttt{bp\_rp} & \textbf{Reddened} $G_{Bp}-G_{Rp}$ colour  & float & Magnitude (mag) \\
\texttt{bp\_g} & \textbf{Reddened} $G_{Bp}- G$ colour  & float & Magnitude (mag) \\
\texttt{g\_rp} & \textbf{Reddened} $G -G_{Rp}$ colour  & float & Magnitude (mag) \\
\texttt{a\_g\_val} & line-of-sight extinction in the $G$ band, $A_G$  & float & Magnitude (mag) \\
\texttt{e\_bp\_min\_rp\_val} & line-of-sight reddening $E(G_{Bp}-G_{Rp})$ & float & Magnitude (mag) \\
\hline
\multicolumn{4}{c}{Spectroscopy}\\
\hline
\texttt{radial\_velocity} & Radial velocity & double & Velocity (km \unit{s}{-1}) \\
\texttt{radial\_velocity\_error} & Standard error of radial velocity \tablenotemark{a} & double & Velocity (km \unit{s}{-1}) \\ 
\hline
\multicolumn{4}{c}{Stellar Parameters\tablenotemark{b}}\\
\hline
\texttt{teff\_val} & Stellar effective temperature & float & Temperature (K) \\
\texttt{lum\_val} & Stellar luminosity &float & Luminosity (Solar Luminosity) \\
\hline\hline
\multicolumn{4}{c}{\textbf{Other fields not in the Gaia DR2 data model}}\\
\hline
\multicolumn{4}{c}{Indices}\\
\hline
 \texttt{parentid} & array \textbf{index} of the generating star particle in the snapshot file & long & \nodata \\
\texttt{partid} & 0 if phase-space coordinates are identical to the generating star particle, 1 otherwise & short & \nodata \\
\hline
\multicolumn{4}{c}{Phase space}\\
\hline
\texttt{ra\_true} & true ra & double & Angle (deg) \\
\texttt{dec\_true} & true dec & double & Angle (deg) \\
\texttt{rad\_true} & true LSR-centric distance & double & Distance (kpc) \\
\texttt{dmod\_true} &  true distance modulus & double & Magnitude (mag) \\
\texttt{parallax\_true} & true parallax & double & Angle (mas) \\
\texttt{pmra\_true} & true pm in ra direction & double & Angular Velocity (mas \unit{yr}{-1}) \\
\texttt{pmdec\_true} & true pm in dec direction & double & Angular Velocity (mas \unit{yr}{-1}) \\
\texttt{radial\_velocity\_true} & true RV & double & km \unit{s}{-1} \\
\texttt{l\_true} & true Galactic long & double & Angle (deg) \\
\texttt{b\_true} & true Galactic lat & double & Angle (deg) \\
\texttt{px\_true, py\_true, pz\_true} & true position relative to LSR\tablenotemark{c}  & double & Distance(kpc) \\
\texttt{vx\_true, vy\_true, vz\_true} & true velocity relative to LSR\tablenotemark{c} & double & km \unit{s}{-1}\\
\hline
\multicolumn{4}{c}{Photometry}\\
\hline
\texttt{phot\_g\_mean\_mag\_true} & \textbf{true} (i.e. after extinction, but before error convolution) apparent $G$-band mean magnitude  & float & Magnitude (mag) \\
\texttt{phot\_bp\_mean\_mag\_true} & \textbf{true} apparent $G_{Bp}$-band mean magnitude  & float & Magnitude (mag) \\
\texttt{phot\_rp\_mean\_mag\_true} & \textbf{true} apparent $G_Rp$ band mean magnitude  & float & Magnitude (mag) \\
\texttt{bp\_rp\_true} & \textbf{true} $G_{Bp}-G_{Rp}$ colour  & float & Magnitude (mag) \\
\texttt{bp\_g\_true} & \textbf{true} $G_{Bp}- G$ colour  & float & Magnitude (mag) \\
\texttt{g\_rp\_true} & \textbf{true} $G -G_{Rp}$ colour  & float & Magnitude (mag) \\
\texttt{phot\_g\_mean\_mag\_int} & \textbf{intrinsic} (i.e. before extinction or error convolution) apparent $G$-band magnitude & float & Magnitude (mag)\\
\texttt{phot\_bp\_mean\_mag\_int} & \textbf{intrinsic}  apparent $G_{Bp}$-band mean magnitude  & float & Magnitude (mag) \\
\texttt{phot\_rp\_mean\_mag\_int} & \textbf{intrinsic}  apparent $G_{Rp}$-band mean magnitude  & float & Magnitude (mag) \\
\texttt{bp\_rp\_int} & \textbf{intrinsic} $G_{Bp}-G_{Rp}$ color  & float & Magnitude (mag) \\
\texttt{bp\_g\_int} & \textbf{intrinsic} $G_{Bp}- G$ color  & float & Magnitude (mag) \\
\texttt{g\_rp\_int} & \textbf{intrinsic} $G -G_{Rp}$ color  & float & Magnitude (mag) \\
\hline
\multicolumn{4}{c}{Extinction}\\
\hline
 \texttt{lognh} & $\log_{10}$ equivalent H column density along line of sight to star & float & \unit{cm}{-2}\\
 \texttt{ebv} & $E(B-V)$ reddening, calculated from \supersub{N}{eff}{H} as discussed in \S \ref{subsec:extinction} & float & Magnitude (mag) \\
 \texttt{A0} & $A_0$, extinction at 550 nm, assuming $R_V=3.1$ (see \S \ref{subsec:extinction}) & float & Magnitude (mag) \\
\hline
\multicolumn{4}{c}{Stellar Parameters}\\
\hline
\texttt{mact} &  current stellar mass & float & Mass (Solar Mass) \\
\texttt{mtip} &  mass of a star at tip of giant branch for given age, metallicity\tablenotemark{d} & float & Mass (Solar Mass) \\
\texttt{mini} &  stellar mass on zero-age main sequence & float & Mass (Solar Mass) \\
\texttt{age} &  log (base 10) of stellar age; identical for all stars generated from the same particle & float & Time (log yr) \\
\texttt{logg} & surface gravity & float & Surface Gravity (log cgs) \\
\hline
\multicolumn{4}{c}{Abundances\tablenotemark{e}}\\
\hline
\texttt{feh} & [Fe/H] & float & Abundances (dex) \\
\texttt{alpha} & [Mg/Fe] & float & Abundances (dex) \\
\texttt{carbon} & [C/H] & float & Abundances (dex) \\
\texttt{helium} & [He/H] & float & Abundances (dex) \\
\texttt{nitrogen} & [N/H] & float & Abundances (dex) \\
\texttt{sulphur} & [S/H] & float & Abundances (dex) \\
\texttt{oxygen} & [O/H] & float & Abundances (dex) \\
\texttt{silicon} & [Si/H] & float & Abundances (dex) \\
\texttt{calcium} & [Ca/H] & float & Abundances (dex) \\
\texttt{magnesium} & [Mg/H] & float & Abundances (dex) \\
\texttt{neon} & [Ne/H] & float & Abundances (dex) \\
\hline
\enddata
\tablenotetext{a}{constant noise floor of 0.11 km \unit{s}{-1} added in quadrature}
\tablenotetext{b}{not error-convolved}
\tablenotetext{c}{see \S\S \ref{subsec:galactocentric}--\ref{subsec:heliocentric} and Table \ref{tbl:solar-positions}}
\tablenotetext{d}{Evolved stars are those with \texttt{mact}$>$\texttt{mtip}.}
\tablenotetext{e}{all relative to solar; see \S \ref{sec:abundances}. Identical for all stars generated from the same particle.}
\end{deluxetable*}
\end{longrotatetable}

\bibliography{refs-all,refs}



\end{document}